\documentclass[prb,aps,amssymb,twocolumn,superscriptaddress,notitlepage]{revtex4-2}
\usepackage{amsmath}
\usepackage{amssymb}
\usepackage{amsthm}
\usepackage{amsfonts}
\usepackage{listings}
\usepackage{physics}
\usepackage{enumerate}
\usepackage{latexsym}
\usepackage{psfrag}
\usepackage{bm}
\usepackage{graphicx}
\usepackage{soul}
\usepackage[caption=false]{subfig}
\usepackage{blkarray}
\usepackage{array}
\usepackage{color}
\usepackage{esint}
\usepackage[normalem]{ulem}
\usepackage{hyperref}
\usepackage[dvipsnames]{xcolor}
\hypersetup{
     colorlinks = true,
     linkcolor = magenta,
     citecolor = magenta
}

\usepackage{comment}
\usepackage{ bbold }
\usepackage{ dsfont }

\def\beq{\begin{equation}}
\def\eeq{\end{equation}}
\def\bal{\begin{aligned}}
\def\eal{\end{aligned}}

\begin{document}
\title{Geometry of Contact Terms in Linear Response: Applications to Elasticity
}
\author{Ian Osborne}
\email{ianro2@illinois.edu}
\affiliation{Department of Physics, University of Illinois at Urbana-Champaign, Urbana IL 61801, USA}
\affiliation{Anthony J.~Leggett Institute for Condensed Matter Theory, University of Illinois at Urbana-Champaign, Urbana IL 61801, USA}

\author{Gustavo M.~Monteiro}
\email{gustavo.monteiro@csi.cuny.edu}
\affiliation{Department of Physics and Astronomy, College of Staten Island, CUNY, Staten Island, NY 10314, USA}

\author{Barry Bradlyn}
\email{bbradlyn@illinois.edu}
\affiliation{Department of Physics, University of Illinois at Urbana-Champaign, Urbana IL 61801, USA}
\affiliation{Anthony J.~Leggett Institute for Condensed Matter Theory, University of Illinois at Urbana-Champaign, Urbana IL 61801, USA}

\date{\today}

\begin{abstract}
Employing the Kubo linear response formalism to calculate the elasticity of anisotropic systems has been shown to yield odd elastic moduli.  For Hamiltonian systems, this result seems to be contradictory as it would violate energy conservation. To resolve this discrepancy, we examine the {predictions}
of quantum linear response in the context of our expectation from classical elasticity theory. 
Our framework reveals that the geometry of the space of strain perturbations introduces correction factors to the correspondence between the Kubo formula and the elastic moduli which resolves the contradiction. 
We use a two-dimensional gas of electrons in a magnetic field as a pedagogical example. 
We use generalized f-sum rules to demonstrate how contact terms may reveal themselves in experimental measurements.
Finally, we discuss the implications of our results for interpreting more general linear response functions.

\end{abstract}
\maketitle

\section{Introduction}

Hydrodynamics and elasticity are enormously successful paradigms for describing the collective behavior of particles, both at the classical and quantum level. Recent progress in particular on the hydrodynamic description of electrons in solids has revealed new phenomena in low-dimensional and strongly correlated systems~\cite{lucas2018hydrodynamics, moll2016evidence,levitov2016electron,bandurin2016negative,pellegrino2017nonlocal,krishna2017superballistic}.
The study of quantum fluids with broken time-reversal symmetry has also brought renewed interest to non-dissipative hydrodynamic response functions such as antisymmetric components of the viscosity tensor, known as Hall (or odd) viscosities, due to their relationship to topological and geometric quantities in gapped systems~\cite{read2009nonabelian,read2011hall,tokatly2006magnetoelasticity,tokatly2007lorentz,levay1995berry,avron1995viscosity,critelli2014anisotropic}. These odd viscosities have since been measured in both quantum systems like graphene~\cite{berdyugin2019measuring} as well as classical active fluids~\cite{soni2019odd}.
On the theoretical side, computation of viscous and elastic response functions has been aided by the development of Kubo type linear response functions for viscosity. While these have been known for some time for time-reversal symmetric systems~\cite{luttinger1964theory,Randeria_Viscosity}, recent extensions to time-reversal broken and anisotropic systems~\cite{bradlyn2012kubo,offertaler2019viscoelastic,rao2020hall} have highlighted the importance of ``diamagnetic'' contact contributions to the viscosity tensor. Building on these works, there has been tremendous progress in developing a geometric interpretation of non-dissipative hydrodynamics in classical and quantum fluids, using a variety of theoretical approaches ranging from Hamiltonian mechanics to quantum field theory~\cite{abanov2019freesurface,abanov2020hydrodynamics,nair2021topological,machado2023hamiltonian,monteiro2024topological,tong2023gauge,monteiro2024korteweg,tong2023gauge}.

While the viscosity in quantum fluids has received the bulk of the attention, the viscosity tensor goes hand-in-hand with other elastic response functions. The viscosity tensor is defined as the response of the stress tensor of a system to velocity gradients (or alternatively, the rate of change of strain); one can consider instead the response of a system to strains directly, which defines the elastic modulus tensor. Conventional wisdom suggests that for fluids, which cannot support shear strains by definition, the elastic modulus tensor should reduce to an isotropic scalar. This is expected even for anisotropic fluids.
In terms of calculations, the full response of the stress tensor to strains can be computed directly from the Kubo formula for viscosity and expanded order-by-order in time derivatives of the strain. In terms of frequency components, this means that the elastic modulus tensor appears in calculations of the viscosity as a simple pole at zero driving frequency. 
Consistent treatment of this pole is essential for disentangling viscous and elastic responses in a calculation. Thus, 
determining the elastic modulus is therefore a crucial component of a consistent viscosity calculation.

Recently, however, it has been shown that the contact terms in the Kubo formula for viscosity can acquire anisotropic and antisymmetric contributions in certain fluids with rotational symmetry breaking. As a particular example, Ref.~\cite{offertaler2019viscoelastic} showed that  breaking the continuous rotational symmetry in a quantum Hall fluid (e.g. by introducing an in-plane magnetic field) induces a finite antisymmetric, component to the diamagnetic term in the Kubo formula, which led those authors to identify a ``Hall elastic modulus.'' In addition to violating our assumptions about stress in fluids, this also raises questions about energy conservation in the strained tilted field system: Under mild assumptions on the dynamics of the Hamiltonian, it can be shown~\cite{scheibner2019odd} that the odd elastic modulus must vanish if energy is conserved.

The apparent emergence of a non-zero odd elastic modulus in anisotropic settings suggests that we must carefully re-examine how the elastic modulus is extracted from the Kubo formula for viscosity. In particular, we must address the question of whether the static elastic response computed via the Kubo formula matches with our classical intuition of how the elastic modulus should be defined. In this work, we will identify the origin of this discrepancy and show how the elastic modulus for a general quantum system can be computed from linear response theory. Our analysis shows that in order to correctly compute elastic response functions, care must be taken to both 1) precisely identify the quantum analogue of classical elastic response, and 2) correctly compose repeated strain transformations to define diamagnetic stresses.

To answer these questions, we will start by reviewing the definitions of elastic quantities, with a focus on how strain is applied to a quantum system at the level of operators and wavefunctions. We show that there are two natural procedures for iteratively straining a system, which yield different expressions for the diamagnetic stress. 
The first is an abelian  process corresponding to the point-wise addition of multiple strain maps acting on a single reference ``equilibrium'' state of the system. 
The second is non-abelian, corresponding to the composition of multiple strain maps.
We argue that while the first procedure is more natural from the point of view of elasticity, it is the second procedure that is most natural in the operator language.
Crucially, the latter procedure is employed by Ref. \cite{bradlyn2012kubo} in their Kubo formulation of elasticity. 
We show that the non-abelian composition of strain maps has a geometric interpretation in terms of Lie groups, and that this precisely accounts for the spurious antisymmetric Hall elasticity of Ref.~\cite{offertaler2019viscoelastic}. 
Importantly, we show that the two different approaches for deriving elastic response {yield equivalent results} as long as the average stress tensor is isotropic in equilibrium. This is true for most fluids, both isotropic and anisotropic (with the quantum Hall fluid in a tilted field being a notable exception). This explains why this issue has gone unnoticed for so long.

While our focus is primarily on contact terms for the elastic modulus, our results highlight a subtlety with broad application to response theory. In particular, by revealing the geometric origin of the corrections to the elastic contact terms,  we argue that similar issues emerge whenever a quantum system is coupled to a set of non-commuting perturbations. Thus, our results have broad applicability. {One natural setting where such geometric considerations appear is in the current-algebra analysis of spontaneously broken chiral symmetries, which leads to the Gell-Mann–Oakes–Renner relation~\cite{dashen1969chiral, WeinbergVol2}. In this framework, the pion mass matrix is fixed by the non-abelian algebra of axial charge densities.} Additionally, in low-energy models of nontrivial flat band systems, the density operators projected into the low-energy subspace fail to commute, instead satisfying analogues of the Girvin-Macdonald-Platzmann algebra; we will argue that for such systems, geometric contributions to contact terms play an important role in studying low-energy response~\cite{Mendez_Valderrama_2024}.
{Finally, the viscoelastic contact term is necessary for the complete description of the non-linear (second order) viscosity which has recently been studied for non-interacting electrons~\cite{jain2025nonlinearoddviscoelasticeffect,jain2025accoustic}.}

This work is organized as follows.
In Section \ref{sec:review}  we review the essential features of classical elasticity theory and constructs a quantum analogue, focusing on the stress and elastic modulus tensors. 
By systematically comparing the different definitions of the elasticity, we arrive at out first key result: the precise definition that we choose for the elastic modulus  has non-trivial physical consequences. 
In Section \ref{sec:Kubo} we present the viscoelastic linear responses within the Kubo formalism and revisit prior calculations of the elastic modulus.  
We identify two distinct parameterizations of strain and show which more closely aligns with classical elasticity theory.
Notably, when the stress tensor is isotropic the 
differences between the two parameterizations serendipitously cancel in the Kubo formula, but in all other cases the methodology developed here is required to arrive at the physically relevant elastic modulus. 
We perform a concrete example using the integer quantum Hall effect, which is of note for possessing an anisotropic stress tensor in the presence of an in-plane magnetic field that breaks the rotational symmetry~\cite{Ian2024}.
Finally, in Section \ref{sec:SumRule} we analyze the elastic modulus from the perspective of linear response sum rules, and propose an experimentally accessible method for measuring the quantum elastic modulus. 

Before proceeding, we will briefly lay out our notational conventions. The Einstein summation convention for repeated indices is implied unless otherwise stated, and tensorial indices are properly oriented to designate either contravariance or covariance.
Operators acting on the Hilbert space carry hats, while all other quantities are understood to be proportional to the Hilbert space identity, $W^i_j \equiv \hat { \mathbb 1}\, W^i_j,$  or to represent expectation values of operators such as $T^i_j = \langle \hat T^i_j\rangle.$ Finally, we will work in units where $\hbar=c=e=1$.

\section{Review of Classical and Quantum Elasticity Theory}
\label{sec:review}

To establish intuition, we begin by first translating the concepts of classical elasticity into the framework of linear response theory. Within this framework, the elastic moduli appear as response coefficients that relate stress to an applied strain. In most experimental situations, one applies the stress and measures the resulting strain. Here, however, we adopt the opposite viewpoint: we assume that the strain can be externally controlled and treat it as the input, while the induced stress is the measured response. From this perspective, the motion of the underlying particles is prescribed by an imposed flow, and the stress is determined by measuring the restoring forces generated in order to maintain that flow. This naturally leads us to interpret the strain as an external field and not as the solution of a dynamical equation. The elastic moduli are then defined as the variations of the stress tensor with respect to the strain, evaluated at elastic equilibrium which is the state of the system in the absence of applied strain.

This setup is analogous to the linear response treatment of electrical conductivity, where the electric field acts as the external perturbation and the induced current is the response. In the present case, the role of the external field is played by the imposed flow that constrains the particle trajectories. To connect this flow-induced strain to more familiar external fields, such as the metric tensor or frame fields, we briefly review the relation between Eulerian and Lagrangian descriptions of continuous media, and how these formulations can be used to describe elastic deformations.

In elasticity, there exists a well-defined equilibrium configuration in which the underlying particles are stationary. For sufficiently small deformations, each particle can be labeled by its equilibrium position. In the continuum limit, these labels identify each particle with a point in space, and the equilibrium configuration defines the material manifold $\mathcal M$, also referred to as the reference manifold. Perfect fluids, by contrast, do not possess a unique equilibrium configuration, since any deformation that preserves the fluid volume costs no energy~\footnote{The presence of shear viscosity introduce friction between the sheared layers of fluid, damping any volume-preserving deformation.}. In this case, the material manifold is instead defined by the initial positions of the particles, which {labels} 
individual fluid elements. In this work, the coordinates on the material space will be denoted by $\{x^i\}$.

\begin{figure}
    \centering
    \includegraphics[width=1\linewidth]{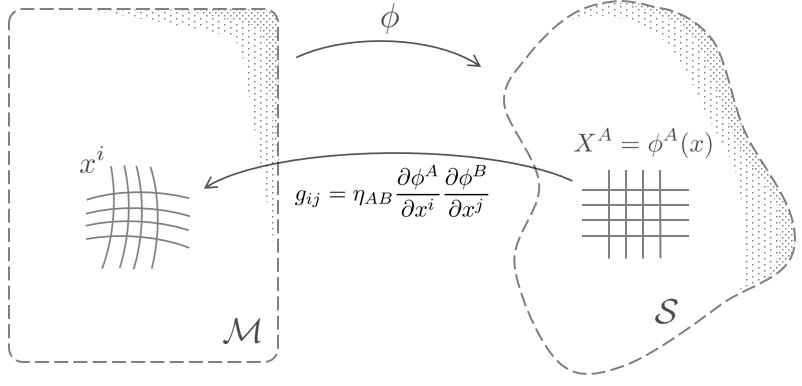}
    \caption{The diffeomorphism $\phi$ from the material manifold $\mathcal M$ to the laboratory (spatial) manifold $\mathcal S$ represents the time-dependent flow of a continuous medium (hydrodynamics) or the action of a dynamical strain applied to an equilibrium configuration (elasticity).}
    \label{fig:1}
\end{figure}

For both perfect fluids and more general elastic media, the particles evolve in time according to a flow, where each particle traces out a trajectory in space, labeled by their material position. 
{Let $\{X^A\}$ denote the particles'} spatial (laboratory) coordinates at time $t$. The trajectory of each particle is then described by
\begin{equation}\label{eq:motion}
X^A = \phi^A(t,x)\,,
\end{equation}
where $\phi$ is the flow map and $X$ is its image. This constitutes the Lagrangian description of the medium. Alternatively, since the collection of particles forms a continuum manifold, $\phi$ may be regarded as a time-dependent deformation of the material manifold $\mathcal M$ into the laboratory  manifold $\mathcal S$ (see Fig \ref{fig:1}), which is often considered to be a subset of the Euclidean manifold $(\mathcal S\subset \mathbb R^d)$. In this coarse-grained description, where particle trajectories are smooth, $\phi$ defines a one-parameter family of diffeomorphisms from $\mathcal M$ to $\mathcal S$, parametrized by the time $t$. Once again, we remind the reader that, in this work, the flow map $\phi$ is an input, not a solution to Newton's equation.

Linear response theory restricts the flow to be a small perturbation around the material configuration. In this regime, the mapping between material and laboratory coordinates can be written as
\begin{equation}\label{eq:Xdef(x,u)}
X^A = \delta^A_i \bigl(x^i + u^i(t,x)\bigr)\,,
\end{equation}
where indices \(i,j,k\) label coordinates on the material manifold \(\mathcal M\), and indices \(A,B,C\) label coordinates on the laboratory manifold \(\mathcal S\).
Because $\mathcal M$ and $\mathcal S$  {are both embedded in $\mathbb R^d$ we may define an identity map which allows for the direct comparison of vectors defined on the two manifolds through the pushforward map $\delta^A_i$.}

This framework allows for a direct comparison between Eulerian and Lagrangian distances. Eulerian distances describe how far apart neighboring particles are in the laboratory frame and are encoded in the laboratory metric tensor $\gamma_{AB}$, which reduces to the identity matrix \(\eta_{AB}\) in Cartesian coordinates~\footnote{In this work, the symbol \(\eta\) denotes the flat Euclidean metric, not the Minkowski metric commonly associated with this notation.}. However, points that are adjacent in the laboratory frame are not necessarily adjacent in the material manifold. As a result, the pullback of the laboratory frame metric to the material manifold {by the strain map $\phi$} induces a time-dependent metric in $\mathcal M$, which differs from the ``natural" (typically Euclidean) metric choice on \(\mathcal M\). Roughly speaking, this induced metric measures separation, within the material space, between particles that lie on neighboring trajectories at time $t$. In field-theory language, \(\mathcal S\) plays the role of the target space and \(\mathcal M\) the role of the base space. Since they are diffeomorphic to one another, the target-space metric induces a metric on the base space via the flow map. Denoting the components of the pullback metric by \(g_{ij}\), we obtain (See Fig. \ref{fig:1})
\begin{equation}
g_{ij} = \frac{\partial \phi^A}{\partial x^i}\,\gamma_{AB}\,\frac{\partial \phi^B}{\partial x^j}\,.
\end{equation}

The laboratory manifold is taken to be a subset of Euclidean space, which allows for the introduction of an orthonormal frame and use the language of Cartan frame fields. In these terms, the induced metric {on $\mathcal M$} takes the form
\begin{equation}\label{eq:metricDef(e,eta)}
g_{ij} = e_i^A \,\eta_{AB}\, e_j^B,
\end{equation}
where the frame field is defined by
\begin{equation}\label{eq:frameDef(phi,x)}
e_i^A \equiv \frac{\partial \phi^A}{\partial x^i}\,
\end{equation}
and its inverse is denoted by $E^i_A = (e^{-1})^i_A$.

In this work, we will focus on time-dependent linear flows, that is,
\begin{equation}
    X^A=\delta^A_i\,\Lambda^i_j(t) \,x^ j=\delta^A_i\,\big(\delta_j^i + \delta\Lambda_j^i(t)\big)\,x^j\,.
\end{equation}
Under this assumption, the displacement is linear 
and the strain field becomes spatially uniform:
\begin{equation} \label{eq:strain-def}
    \frac{\partial u^i}{\partial x^j}=\delta\Lambda^i_j(t).
\end{equation}

{Using equations \eqref{eq:motion}, \eqref{eq:Xdef(x,u)}, and \eqref{eq:frameDef(phi,x)},  the Cartan frame field and the strain tensor are related by a change of index.
\begin{align}\label{eq:eLambdaID}
    e_j^A = \delta^A_i \Lambda^i_j
\end{align}
It is important to note that while the analogy is useful, there are essential distinctions between general relativity and elasticity. For instance, the frame fields used in this work do not possess an ``internal" symmetry associated with local rotations in the lab frame: $e_i^A \rightarrow R^A_B\, e^B_i$. The lab frame possesses uniform, spatially independent orthogonal basis vectors which are extrinsically imposed by the setting of the lab.}

{
Due to the linear assumption made of $\Lambda$ and by analogy of equation \eqref{eq:metricDef(e,eta)}, the induced metric is independent of the material coordinate:}
\begin{align}
g_{ij}
= e_i^A e_j^B\, \eta_{AB}
= \Lambda_i^k(t) \,\Lambda_j^l(t)\,\eta'_{kl}\,.
\end{align}
Where $\eta'_{kl} $ is a Euclidean metric on $\mathcal M$.

Notice that \(\Lambda^k_i\) carries only material indices, whereas \(e_i^A\) is a genuine two-point tensor relating \(\mathcal M\) to \(\mathcal S\). From this perspective, the matrix $\Lambda^k_i$ describes a diffeomorphism from the material manifold \(\mathcal M\) to itself, mapping its Euclidean metric into the induced metric $g_{ij}(t)$, which encodes information about the laboratory frame $\mathcal S$:
\begin{equation}
    x'{}^i=\Lambda^i_j \,x^j\quad \text{and}\quad X^A=\delta^A_i\,x'{}^i.
\end{equation}
The advantage of this formulation is that {it makes a direct connection to classical elasticity, where $\delta\Lambda^i_j$ correspond to the strain field $\partial_ju^i$. In addition to that,} all raising and lowering of indices can be performed using the time-dependent metric \(g_{ij}\) and its inverse \(g^{ij}\), without explicitly converting between the laboratory manifold \(\mathcal S\) and the material manifold \(\mathcal M\), as raising indices with $g^{ij}$ is mechanically identical to mapping tensors from $\mathcal M $ to $\mathcal S$, raising the index with the Euclidean $\eta^{AB}$, and then mapping back to $\mathcal M$. 

{To understand how strains are encoded in quantum mechanics, we examine the expectation value of strained operators. Strains can be treated either as active transformations on states or as passive transformations on operators, analogous to the Heisenberg picture. In the latter viewpoint, operators in the lab frame are the strained operators, which can be obtained from the material frame operators through a similarity transformation:
\begin{align}\label{eq:OperatorStrain}
    \hat A_\Lambda = \hat S \hat A \hat S^{\dagger},
\end{align}
where $\hat S$ is the unitary operator associated with constant strains. By definition, the expectation value of a one-body operator in the material frame is
\begin{align}\label{eq:A-MaterialFrame}
    \langle \hat A\,\rangle
    = \int_{\mathcal M} d^d x \; \psi^\dagger(x)\,
     \hat A \, \psi(x).
\end{align}
However, the expectation value evaluated directly in the lab frame automatically incorporates the effect of strain. Expressing the expectation value of the same operator {using} material coordinates, yields
\begin{align}\label{eq:A-LabFrame}
    \langle \hat A_\Lambda\rangle
    = \int_{\mathcal M} d^d x \; \mathcal J\,
      \psi^\dagger(\Lambda x)\, \hat S \hat A \hat S^{\dagger}\, \psi(\Lambda x),
\end{align}
where $\mathcal J = \det(\Lambda)$ is the Jacobian of the coordinate transformation. By comparing Eqs.~\eqref{eq:A-MaterialFrame} and \eqref{eq:A-LabFrame}, we see that the strain operator $\hat S$ must satisfy
\begin{align}\label{eq:SDEF}
   \hat S\psi(x,t)
   \equiv \sqrt{\mathcal J}\,\psi(\Lambda x,t)
   = \exp\!\left(\frac{i}{2}\lambda^{i}_{j}
     \{\hat x^{j},\hat p_{i}\}\right)\psi(x,t)\,,
\end{align}
where $\hat p_i = - i \partial_i$ is the (canonical) momentum operator. In deriving this, we assumed the absence of magnetic field and used
\begin{align}\label{eq:framelambda}
    \Lambda^{i}_{j} = \big(\exp \lambda\big)^{i}_{j},
    \qquad
    \det(\Lambda) = e^{\mathrm{tr}(\lambda)}.
\end{align}
The anticommutator in Eq.~\eqref{eq:SDEF} guarantees unitarity and ensures that the wavefunction transforms correctly as a scalar density of weight $1/2$, as reflected in the factor of the Jacobian.}

For multiple particles, the generator of  constant strains is  defined by straining the coordinates of every particle, yielding the strain operator (Eq. \eqref{eq:SDEF}).
\begin{align}\label{eq:StrainOpp}
    \hat S = \exp(-i \lambda_i^j \hat{J}^i_j)\,,\\
   \hat  J _j^i = - \frac{1}{2}\sum_n\{\hat x^i_n, \hat p_j^n\}\,, \label{eq:StrainGen}
\end{align}
where $n$ is a particle index. In the absence of external magnetic fields $\hat J$ acts on position and momentum operators by 
\begin{align}\label{eq:JPXcom}
    [ \hat{J}^i_j,\hat p_k] = -i \delta_k^i \hat p_j,~~~~~[ \hat{J}^i_j,\hat x^k] =i \delta^k_j\hat  x^i.
\end{align}
The strain generators are closed under commutation.
\begin{align}\label{eq:LieAlg}
 [\hat J_{j}^i,\hat{J}_l^k] = i \delta_{j}^k\hat J_{l}^i - i \delta_l^i\hat J_j^k
\end{align}

In the presence of magnetic fields, the strain generators also generate {two additional transformations. The first is a gauge transformation that is necessary to maintain Eq.~\eqref{eq:LieAlg}. The second is} a rescaling of the magnetic field  due to the definition of the magnetic field as $B^i = \epsilon^{ijk}\partial_j A_k$ and the Levi-Civita tensor which is $\epsilon^{ijk} = \widetilde \epsilon^{\,ijk}/\sqrt{g}$ where $\widetilde \epsilon^{123} = 1$ and is perfectly antisymmetric. This preserves the total flux through the system instead of the field strength.  Ref. \cite{bradlyn2012kubo} demonstrates a method for modifying the strain generator such that equations \eqref{eq:JPXcom} and \eqref{eq:LieAlg} remain true even for nonzero magnetic fields, and as a consequence, the magnetic field in the material frame is a constant throughout a strain transformation. 

{The lab Hamiltonian projected back to the reference manifold is 
\begin{align}\label{eq:perturbation}
    &\hat H_{\Lambda} = \hat S \hat H \hat S^{\dag}= \hat H + \hat H' \,, \\
    \label{eq:Hprime}
    &\hat H' = \sum_{n=1}\frac{(- i)^n[\lambda_i^j \hat J_i^j,\hat H]_n}{n!}\,,\\
    &[\hat A,\hat B]_n = [\hat A,[\hat A,\overset{\text{ n}}{...},[\hat A,\hat B]]]\,.
\end{align}
Notice that this is expanded in powers of the matrix $\lambda$, however, which differs from the strain $\delta\Lambda$ in Eq.~(\ref{eq:strain-def}). The two do coincide to linear order expansion, but not at the second order. In terms of $\delta\Lambda$, the Hamiltonian expansion becomes:
\begin{align}\label{eq:perturbedHam}
     \hat H_\Lambda&=\hat H-i \delta\Lambda_i^j [\hat{J}^i_j, \hat H] \\
     &+ \frac{1}{2} \delta \Lambda_{i}^j \delta \Lambda_k^l\left(-[\hat{J}^i_j ,[\hat J_{l}^k,\hat H]] + i\delta^k_j  [\hat{J}^i_l,\hat H]\right)+O(\delta\Lambda^3)\,. \nonumber
\end{align}}

{Here, we used that
\begin{align} \label{eq:lambda-ln(Lambda)}
    \lambda=\ln(\mathbb 1+\delta\Lambda)=\sum_{n=1}^\infty(-1)^{n+1}\frac{(\delta\Lambda)^n}{n}\,.
\end{align}}
{Eq.~\eqref{eq:perturbedHam} differs at quadratic order from the expansion in terms of $\lambda$ used in previous works~\cite{bradlyn2012kubo,Hughes_CSMM,link2018elastic,offertaler2019viscoelastic,rao2020hall}}.
{ Even though the expansions in $\lambda$ and in $\delta\Lambda$ coincide at the linear order, a clear distinction arises when studying the elastic response, whose leading contribution comes from the quadratic term. The difference between taking derivatives with respect to $\lambda$ versus $\Lambda$ can be computed using the chain rule and Eq.~\eqref{eq:lambda-ln(Lambda)}, and is given by
{
\begin{align}
    \frac{\partial }{\partial\Lambda_{l}^k}  \bigg|_{\Lambda = \delta}&=   \frac{\partial }{\partial\lambda_{l}^k}  \bigg|_{\lambda = 0}\\
    \begin{split} \frac{\partial^2 }{\partial\Lambda_{l}^k\partial\Lambda^i_j} &\bigg|_{\Lambda = \delta}\\
     =   \bigg( \frac{\partial ^2}{\partial\lambda_{l}^k\partial \lambda_j^i}&- \frac{1}{2}\left(\Gamma^{\binom{n}{m}}  _{\binom{j}{i}\binom{l}{k}}+ \Gamma^{\binom{n}{m}}  _{\binom{l}{k}\binom{j}{i}}\right)\frac{\partial}{\partial \lambda^n_m}\bigg) \bigg|_{\lambda = 0},\label{eq:ChainRule}
     \end{split}
\end{align}
where 
\begin{equation}\label{eq:Gamma}
    \Gamma^{\binom{n}{m}}  _{\binom{j}{i}\binom{l}{k}}\bigg|_{\lambda = 0} =   \delta^l_i \delta^j_m \delta^n_k .
\end{equation} 
Our choice of definition for $\Gamma$ in Eqs.~\eqref{eq:ChainRule} and \eqref{eq:Gamma} will be further motivated in sections \ref{sec:contact term comparison} and \ref{sec:delta g T}.
The expansion Eq.~\eqref{eq:perturbedHam} and the expression for the second derivatives with respect to $\lambda$ or $\delta\Lambda$ is the first result of this work. }

{At this point, it is convenient to trace a parallel to standard classical elasticity. In contrast to the picture described so far, the strain dynamics are not fixed in classical elasticity, but determined by the elastic forces in the system. These restoring forces are obtained through an elastic potential, under the assumption that the system is restricted to small perturbations on the equilibrium configuration. In general, the elastic potential energy can be written as
\begin{equation}
    U=\frac{1}{2}\int_{\mathcal M} d^dx \,A^i{}_j{}^k{}_l\,\frac{\partial u^j}{\partial x^i}\frac{\partial u^l}{\partial x^k}\,.
\end{equation}
Here, the elasticity tensor $A^i{}_j{}^k{}_l$ is symmetric under the swap between the first two indices and the last two, that is,} { $(ij)\leftrightarrow (kl).$ }
{This invariance will be referred to as the \textit{hyper-elastic symmetry}~\cite{Marsden_Elasticity} and it relies on an underlying Hamiltonian structure.

{Along with the elasticity, the stress tensor encodes information of the local forces acting on the fluid (see appendix \ref{sec:Review}), and hence also conveys details about the work needed to deform the system.} For strains defined in Eq.~(\ref{eq:strain-def}), the stress tensor takes the form
\begin{align}
    T^{i}_{j}&= -\frac{\partial U}{\partial \Lambda^{j}_{i}}\,.
\end{align}
A reasonable extension to quantum mechanical systems is obtained by making the substitution $U \rightarrow \langle H_\Lambda \rangle$. In addition, since the Jacobian $\mathcal J = \det(\Lambda)$ encodes the strain-induced change in volume, it is therefore natural to define the expectation value of the quantum mechanical stress tensor as
\begin{align}\label{eq:stress-from-classical}
    \langle\, \hat T^{i}_{j}\,\rangle
    &\equiv -\frac{1}{\mathcal J}
      \frac{\partial \langle \hat{H}_\Lambda\rangle}
           {\partial \Lambda^{j}_{i}}\,.
\end{align}}

{This discussion emphasizes two central points that motivate the present work. First, the elastic tensor of a Hamiltonian quantum system must exhibit the hyper-elastic symmetry. Second, the introduction of the transformation $\Lambda^{i}_{j}$ provides a bridge between the operator formulation in quantum mechanics and the description of the system through an induced action in a geometric background.}

{
Due to the fact that the strain transformation relates two different manifolds $\mathcal{M}$ and $\mathcal{S}$, one must choose which manifold in which to measure these local forces, naturally resulting in differing definitions. 
This is an often overlooked detail because the quantitative difference between any two method of calculating the stress vanishes in the limit of small strain.
The same is not true for the derivative of the stress with respect to the strain: elastic moduli, which possess finite differences depending on which method is chosen. 
We will briefly review the main methods for defining stress on the two manifolds. 
}

For systems with hyper-elastic symmetry (hyper-elastic systems),  the first Piola-Kirchhoff (P-K) stress~\cite{Marsden_Elasticity} is  defined as the derivative of the elastic potential with respect to the frame fields. Quantum mechanically, we can write
\begin{align}\label{eq:FirstPiola}
      P^{i}_A =-\frac{1}{\mathcal J} \frac{\partial \langle \hat H_{\Lambda}\rangle}{\partial e_i^A} 
\end{align}

Because the frame fields in our setup are related to $\Lambda$ by equation \eqref{eq:eLambdaID}, $P$ is the quantity most related to the stress calculated in equation \eqref{eq:stress-from-classical},
\begin{align}\label{eq:FirstPiola2}
    T^i_j\equiv  -\frac{1}{\cal{J}}\frac{\partial\langle\hat{H}_\lambda\rangle}{\partial e_k^A}\frac{\partial e_k^A}{\partial \Lambda^j_i}=\delta_j^A P^i_A 
\end{align}
Despite this, the two are mathematically distinct since $P^i_A$ is a two-point tensor with one index in $\mathcal M$ and the other in $\mathcal S$. Its main utility is the calculation of the local forces in the lab frame by taking a divergence in the material frame {where the coordinates are independent of the strain}: $f_A^{\text{local}} = - \nabla_i p^i_A$ where $p$ is the intensive form of $P$, {and $\nabla$ is the covariant derivative under the induced metric.}
The second P-K stress is the tensor with all indices on the reference manifold,
\begin{align}\label{eq:SecondPiola}
    \mathcal T^{ij}\equiv E^{j}_A P^{iA}. 
\end{align}

In many fluids, including {all those with rotational symmetry (and some without rotational symmetry)}, the elastic potential depends only on the induced metric (i.e. only on distances), and not on the frames directly. In this case, the second P-K stress can be written directly as a derivative of the elastic energy via
\begin{equation}
\mathcal T^{ij}= - \frac{2}{\mathcal J} \frac{\partial \langle \hat H_{\Lambda}\rangle}{\partial g_{ij}}.
\end{equation}

The lab frame counterpart of the second P-K stress, the Cauchy stress tensor $\sigma^{AB}$, is likely the most familiar form of the stress tensor and both indices point in $\mathcal S$. As such, the Cauchy stress conveys information about forces acting on a small box of fluid in the lab frame. The Cauchy stress is given by
\begin{align}\label{eq:Cauchy}
    \sigma^{AB} \equiv \mathcal J e^B_i \eta^{CA}P^i_C=- \frac{2}{\sqrt{\det \gamma}} \frac{\partial \langle \hat H\rangle}{\partial\,\gamma_{AB}}\bigg|_{\gamma = \eta},
\end{align}
where the second equality again is valid when the elastic energy depends only on distances, and generalizes to Lorentz invariant systems as the spatial part of the stress-energy tensor~\cite{Brown_2021}.

Finally, the last kind of stress we will mention is the kind calculated in Ref. \cite{bradlyn2012kubo,offertaler2019viscoelastic} by the Ward identity which is the statement that around $\delta \Lambda = 0, $ the stress tensor operator is given by 
\begin{align}\label{eq:Ward}
     (\hat T_0)^i_j = i [\hat{J}^i_j,\hat H] = -\frac{\partial }{\partial \lambda_i^j} \hat H'.
\end{align}
This can be seen from equations \eqref{eq:perturbation} and \eqref{eq:Hprime}.
Because $\lambda = \delta \Lambda + \mathcal O(\delta\Lambda^2)$, the Ward identity gives the stress operator evaluated in the unstrained system.

Each of the stress tensors mentioned above coincide quantitatively at equilibrium, but they differ when evaluated at nonzero strain.
Therefore, they each possess different linear expansions in terms of strain parameters. {As the elasticity tensor is defined through the linear expansion of the stress in terms of strain, the different stresses lead to inequivalent elasticity tensors.
Unlike the relatively minor differences in the definitions of the stress tensors, the components of the different elasticity tensors differ by nonzero correction terms even in the flat background.}

Let's quickly discuss the two most relevant elasticity tensors.
The first, $A$, is the (negative) expansion of the first P-K stress in terms of the frame field.
\begin{align}\label{eq:firstElasticity}
    A^{i~j}_{A~B} = - \frac{1}{\mathcal J}\frac{\partial \mathcal J P^i_A }{\partial e_j^B}= - \frac{1}{\mathcal J}\frac{\partial \mathcal J P^j_B }{\partial e_i^A}
\end{align}

The first elasticity possesses the hyper-elastic symmetry $(iA)\leftrightarrow (kB)$.  Note that the first elasticity can be defined for any system regardless of anisotropy. {Moreover, it is straightforward to verify that this expression is directly connected to the classical elastic tensor via 
{
\begin{align}\label{eq:Adeltadelta}
    A^{i}{}_{k}{}^{j}{}_{l}=A^{i~j}_{A~B}\,\delta^A_k\delta^B_l\,,
\end{align}}}{which projects the lab indices onto material indices. It is worth emphasizing that this projection is performed using the Kronecker delta, rather than via the frame fields, {due to our interpretation of the lab and material frames as both embedded in the same Euclidean space}.}

The second elasticity is defined by a derivative with respect to the metric, and so can only be defined for systems where the elastic energy depends only on the distance between particles. It is given by
\begin{align}\label{eq:secondElasticity}
    K^{ijkl} = - \frac{1}{\mathcal J} \frac{\partial \mathcal J \mathcal T^{ij}}{\partial g_{kl}} = - \frac{1}{\mathcal J} \frac{\partial \mathcal J\mathcal T^{kl}}{\partial g_{ij}}. 
\end{align}

The elastic tensor $K$ possesses the hyper-elastic symmetry of the first elasticity $(ij)\leftrightarrow (kl)$. It also inherits additional symmetries from the metric, namely the exchange symmetries $i\leftrightarrow j$ and $k\leftrightarrow l$.
In classical elasticity, it is almost exclusively the second elasticity that is used~\cite{Marsden_Elasticity,tokatly2006magnetoelasticity,folkner2024elasticmodulithermalconductivity,deJong2015,ElasticityPathIntegral2001} along with its analog in the material frame,
\begin{align}
    C^{ABCD} = -e_i^Ae_j^Be_k^Ce_l^D K^{ijkl}  =-\frac{1}{\sqrt{\gamma }}\frac{\partial\,\sqrt{\gamma}\sigma^{AB}}{\partial\,\gamma_{CD}}\bigg|_{\gamma = \eta}.
\end{align}
The components of $C$ coincide with those of $K^{ijkl}$ when evaluated in the unstrained configuration $e = \mathbb 1$.

A key subtlety going forward is the relationship between $A$ and $K$. Even though $\mathcal T^{ij}$ is related to $P^{iA}$ by contraction with the inverse frame field $E_A^j$, the elastic constant $K$ is not as simply related to $A$. This is because the inverse frame field has a nontrivial derivative with respect to other frame fields. Taking care of this subtlety, we find that $A$ is related to $K$ via
\begin{align}\label{eq:AKconversion}
    A^{iAkB} &= -\frac{\eta^{BC}}{\sqrt{g}}\frac{\partial \sqrt{g} e_j^A \mathcal T^{ij}}{\partial e_k^C}\\
     &= \frac{e_j^A K^{ijnm}}{\sqrt{g}}(e_{n}^B \delta_m^j + e_{m}^B\delta ^j_n) -\mathcal T^{ik} \eta^{AB}.\label{eq:secondMod}
\end{align}
Because the elasticity tensors are given by second derivatives of $\langle \hat H_\Lambda\rangle$ (i.e. because they are susceptibilities), we must use Kubo linear response to evaluate them for quantum systems. This will be the main goal of Sec.~\ref{sec:Kubo}.

Note that the second P-K stress and the Cauchy stress are not well-defined as derivatives of the elastic potential with respect to the spatial metric if there are other rank-2 tensors which enter the Hamiltonian. For general condensed matter systems, lattice effects can lead to anisotropy where, e.g. the band mass tensor or dielectric tensor are not proportional to the spatial metric.  Therefore in anisotropic systems, the second P-K and Cauchy stresses must be defined by equation \eqref{eq:SecondPiola} or the first equality in equation \eqref{eq:Cauchy}. {The elastic modulus $K$ is undefined for these systems as its definition} requires the strain perturbation to enter in the Hamiltonian only by factors of $\delta g_{ij} = \eta'_{ik}\delta \Lambda^k_j + \delta\Lambda^k_i \eta'_{kj}$.
Equation \eqref{eq:secondMod} cannot be used to generalize the second elasticity (see appendix \ref{sec:anisotropy} for details), the first elasticity $A$, however is always definable.

Finally, let us comment on the implications of the Ward identity [Eq. \eqref{eq:Ward}] for the elastic constants. As we have stated, the Ward identity only calculates the stress at equilibrium to zeroth order in the strain perturbation, so expanding $(T_0)^i_j = \langle(\hat T_0)^i_j\rangle $ in terms of $\lambda$ yields an elasticity which is distinct from both $A$ and $K$. 
The analogous elasticity computed by taking second derivative with respect to $\lambda$ has an attractive geometric and physical interpretation which we discus further in section \ref{sec:contact term comparison}. It is also this quantity that appears in previous Kubo calculations of the viscosity and elasticity. However, as we will explore further in Sec.~\ref{sec:contact term comparison}, Eq.~\eqref{eq:ChainRule} implies that the elastic constant computed using second derivatives with respect to $\lambda$ is distinct from both $A$ and $K$ in general.

In the next section, we look at the quantum mechanical calculation of the viscoelastic linear response from the Kubo formula, and show how to incorporate these insights to compute the first- and second elasticity tensors from a revised formulation. We will show that $\Gamma$ from Eq.~\eqref{eq:Gamma} reflects the nontrivial geometry of the space of strain perturbations.

\section{Elastic Modulus from Kubo}
\label{sec:Kubo}

In this section, we derive the viscoelastic Kubo formulas and discuss the new corrections due to the non-trivial geometry of the space of strains. 
For simplicity, we will work in the material frame from here forward by converting all indices to material indices and all derivatives with respect to $e_i^A$ to derivatives of $\Lambda_i^j$ {via equation \eqref{eq:eLambdaID}.}

Let $\hat H$ be an arbitrary unstrained Hamiltonian, and let the strain perturbation be given by equation \eqref{eq:Hprime}.
The Kubo formula expresses the linear response of the stress tensor to a strain perturbation as  
\begin{align}\label{eq:Kubo}\begin{split}
    &\langle  \hat T^i_j (t)\rangle  - \langle \hat T^i_j\rangle_0\\&= \left\langle (\delta_\Lambda \hat T)^{i~k}_{~j~l}\right\rangle_0\, \delta \Lambda^l_k(t) + \int dt' (\chi^{TT})^{i~k}_{~j~l}(t- t') \delta\Lambda_k^l(t'), 
    \end{split}
\end{align}
where we have defined
\begin{align}\label{eq:deltaT}
    (\delta_\Lambda \hat T)^{i~k}_{~j~l} ~\equiv  \frac{\partial \hat  T_j^i(0)}{\partial  \Lambda_k^l},
\end{align}
and $\langle \dots\rangle _0 $ is the expectation in the unperturbed ground state.
The first term of equation \eqref{eq:deltaT} is a contact term which, in the analogous calculation of the conductivity, is known as the diamagnetic term. Its appearance reflects the fact that the definition of the stress operator depends on the value of the strain field~\cite{ROSTAMI2021168523}. 
The second term is the stress-stress correlated response term given by 
\begin{align}\label{eq:susceptibility}
    \chi^{TT}_{ijkl}(t) = -i\Theta(t) \langle[\hat T_i^j(t), \hat T_k^l(0)] \rangle_0,
\end{align}
where  the time evolution of $\hat T_i^j(t)$ is determined by the unperturbed Hamiltonian $\hat H$. 
We will refer to this as the integral term of the Kubo formula.
For viscous systems, the integral term is crucial as the viscosity is the element of the stress generated by time derivatives of the strain, $\delta T^i_j = \eta ^{i~k}_{~j~l} \partial _t \Lambda_k^l$.
In equilibrium, we may use the Ward identity [Eq. \eqref{eq:Ward}] to calculate the operator $\hat T$ as all terms linear order in $\delta \Lambda$ are set to zero.

The same is not true for the diamagnetic term which is explicitly the linear order expansion of the stress operator.
The diamagnetic term in equation \eqref{eq:Kubo} is \textbf{not} the first elasticity, but rather 
\begin{align}\label{eq:omegais0}
    A^{i~k}_{~j~l} = -\left\langle  (\delta_\Lambda \hat T)^{i~k}_{~j~l} \right\rangle_0  - \int dt' (\chi^{TT})^{i~k}_{~j~l}(t')
\end{align}
since the first elasticity also incorporates the $\omega = 0$ contribution to the susceptibility.
{Because we define $A$ with all material indices by equation \eqref{eq:Adeltadelta}, the indices $j$ and $l$ are effectively lab indices referring to directions in $\cal{S}$, and should be raised and lowered with the lab frame metric $\eta'_{ij}= \delta^A_i \delta^B_j\eta_{AB}$; whereas $i$ and $k$ are material indices referring to directions in $\cal{M}$ and must be raised and lowered by the induced metric $g_{ij}$:} 
\begin{align}
      A_{ijkl} &= g_{in}g_{km}A^{n~m}_{~j~l}{=g_{in}g_{km}\delta^A_j\delta^B_lA^{n~m}_{~A~B}\,}, \\
      A^{ijkl} &= \eta'^{jn}\eta'^{lm}A^{i~k}_{~j~l}{=\delta^j_A\delta^l_B\eta^{AC}\eta^{BD}A^{i~k}_{~B~C}}.
\end{align}
This subtlety may be ignored only at the elastic equilibrium point when the matrix elements of either metrics are identical $\eta = g = \delta$.

\subsection{Prior Formulation of the Elastic Contact Term}\label{sec:contact term comparison}

In previous works~\cite{offertaler2019viscoelastic,bradlyn2012kubo,rao2020hall}, $\lambda$ (instead of $\delta \Lambda$) was used as the strain perturbation. From the Hamiltonian perspective, the expansion of $\hat H$ to second order in $\lambda$ produces an elastic diamagnetic term:
\begin{align}\label{eq:partial2Hpartiallambda}
 \frac{\partial^2 \hat H_\Lambda}{\partial \lambda_j^i \partial\lambda_k^l} =  \frac{1}{2}( [\hat J^i_j,[\hat J^k_l,\hat H]]+[\hat J^k_l,[\hat J^i_j,\hat H]]  ).
\end{align}
We have arrived at the above equation by symmetrizing the quadratic term in equation \eqref{eq:perturbation}.
While explicit symmetry of the quadratic term under $(ij)\leftrightarrow (kl)$ is not required, any antisymmetric component will sum to zero when contracted with $\lambda_i^j\lambda_k^l$.

From the perspective of Kubo linear response, the elasticity is given by expanding the stress in terms of strain perturbations. Since Eq.~\eqref{eq:OperatorStrain}, gives the change in the stress tensor under a strain transformation, it is a natural starting point for computing the diamagnetic stress.
Previous works on the Kubo formula for viscoelastic response in quantum mechanics have thus used
\begin{align}\label{eq:OffertalerK}
 ( \Delta_\lambda \hat T_0)^{i~k}_{~j~l} \equiv \frac{\partial}{\partial \lambda_k^l} \hat S (\hat T_0)^{i}_j\hat S^{\dag}= -i [\hat{J}_l^k, (\hat T_0)_j^i] = [\hat J_l^k,[\hat J^i_j,\hat H]]
\end{align}
for the elastic contact term~\cite{bradlyn2012kubo,offertaler2019viscoelastic}. We denote the expression shown in equation \eqref{eq:OffertalerK} as $\Delta_\lambda \hat T_0$ to distinguish it from the physical contact term $\delta_\Lambda \hat T$. As we will see, $\Delta_\lambda \hat T_0 $ does not correspond to the partial derivative in equation \eqref{eq:deltaT}.

First, note that equation \eqref{eq:OffertalerK} does not coincide with the contact term found in equation \eqref{eq:partial2Hpartiallambda} when the commutator $[\hat J^k_l,(\hat T_0)^i_j]$ is not symmetric about $(ij)\leftrightarrow(kl)$. 
We will soon discuss the conditions where this is the case on average in the ground state.  Next, compare equation \eqref{eq:partial2Hpartiallambda} with the expansion of $\hat H_\Lambda$ in terms of $\delta \Lambda$ we found in equation \eqref{eq:perturbedHam}. We find that the manually symmetrized contact term is not equal to the physical contact term implying the choice to expand in terms of $\delta \Lambda$ versus $\lambda$ is physically meaningful.

Equation \eqref{eq:OffertalerK} is analogous to derivatives of the electromagnetic gauge field which can be identified with commutators of the charge density.
\begin{align}\label{eq:PartialA}
   q_i \frac{\delta \hat O}{\delta A_i} = -i [\hat \rho_q, \hat O ] 
\end{align}

As hinted, $\Delta_\lambda \hat T_0 \neq \partial \hat T_0/\partial \lambda$ is \textbf{not} a partial derivatives with respect to $\lambda $ rather the operation in equation \eqref{eq:OffertalerK} represents a Lie
 derivative on the space of strains $GL(d,\mathbb R) $
The direction of the Lie derivative 
is set by the infinitesimal strain $\lambda$, and can be written as $\lambda^j_i\overline v^i_j$ where $\{\overline v^i_j\}$ is a set of basis vectors at the identity element in the Lie algebra $\mathfrak {gl}(d,\mathbb R)$ of equation \eqref{eq:LieAlg} and is represented in the Hilbert space by  $\hat{J}^i_j$. Specifically, we have 
\begin{align}\label{eq:LieDeriv}
   \mathcal L_{
   \lambda_i^j\overline v^i_j} \mathcal O &=-i[\lambda_i^j\hat{J}^i_j,\hat O]  ,\\
     \mathcal L_{
     \lambda_i^j\overline v^i_j} \mathcal  O&\equiv \lim_{t \rightarrow 0}\frac{1}{t}\left( \mathcal O_{\exp(t\lambda)  }
    -\mathcal O_{\mathbb 1}\right),
\end{align}
where $\hat O$ is the representation of the Lie group field $\mathcal O$ on the Hilbert space.
Crucially, variational derivatives and Lie derivatives only coincide  when the Lie group is abelian as in the $U(1)$ vector potential example [Eq. \eqref{eq:PartialA}] or when applied to scalar fields  such as the Hamiltonian.
Reinterpreting the Ward identity \eqref{eq:Ward}, we can write the stress tensor operator in equilibrium as
\begin{align}\label{eq:SpencerDeriv}
    (\hat T_0)^i_j\lambda_i^j = - 
    \lambda_i^j \overline{v}^i_j( \hat H),
\end{align}
which implies it is a covector on the manifold of strains. This explains why $\hat T_0$ does not transform under strain in the same manner that scalar quantities such as the Hamiltonian transform, i.e., by equation \eqref{eq:OperatorStrain}.

The application of a second Lie derivative  gives $\Delta_\lambda \hat T_0$ in equation \eqref{eq:OffertalerK},
\begin{align}
   - \mathcal L_{(\lambda_2)^l_k\overline v^k_l} (\mathcal L_{(\lambda_1)^j_i\overline v^i_j}\hat H)&=(\lambda_1)^j_i\mathcal L_{(\lambda_2)^l_k\overline v^k_l} (\hat T_0)^i_j \nonumber
   \\ 
   &= (\lambda_2)^l_k( \lambda_1)^j_i[\hat{J}^k_l,[\hat{J}^i_j,\hat H]] \nonumber \\
   &\equiv (\Delta_\lambda \hat T_0)^{i~k}_{~j~l}(\lambda_2)^l_k( \lambda_1)^j_i. \label{eq:D2}
\end{align}
The resulting tensor is not symmetric about $(ij)\leftrightarrow(kl)$ which would be required by the commutativity of the partial derivatives in equation \eqref{eq:firstElasticity}.
Lie derivatives have no such symmetry for general Lie groups. In fact, the odd/antisymmetric part of Eq.~\eqref{eq:D2} is related to the structure constants of the Lie algebra via the identity
\begin{align}\label{eq:LieIdentity}
    \mathcal L_{\overline v_a} \mathcal L_{\overline v_b } - \mathcal L_{\overline v_b} \mathcal L_{\overline v_a }= \mathcal L_{[\overline v_a,\overline v_b]}.
\end{align}
For convenience we have bundled the $(ij)$ indices into a single index $a.$ 
In terms of commutators with $J$, this is equivalent to the Jacobi identity. The antisymmetric component of $(\Delta_\lambda\hat T_0)^{i~k}_{~j~l}$ for an arbitrary Hamiltonian is
\begin{align}
\begin{split}\label{eq:Jacobi}
     &(\Delta_\lambda\hat T_0)^{i~k}_{~j~l}- (\Delta_\lambda\hat T_0)^{k~i}_{~l~j}=   [ \hat J^{k}_l,[ \hat J^{i}_j,\hat H ]]+  [ \hat J^{i}_j,[\hat H, \hat J^{k}_l]],\\
     &= - [\hat H ,[\hat  J^{k}_l, \hat J^{i}_j]]
     = -\delta^{k}_j(\hat T_0)_l^{i} + \delta^i_l(\hat T_0)^{k}_j,
     \end{split}
\end{align}
where we have used the commutation relations in equation \eqref{eq:LieAlg}.

For the purposes of the Kubo response formula $\chi^{TT}_{ijkl}(\omega) = \chi^{TT}_{klij}(-\omega)$, and since the susceptibility enters the calculation of the elasticity only through the $\omega = 0$ component, the integral term in equation \eqref{eq:omegais0} possesses the hyper-elastic symmetry. Thus, the hyper-elastic symmetry is broken by the contact term [Eq. \eqref{eq:Jacobi}] whenever $\langle (\hat T_0)_i^j\rangle\neq P \delta_i^j $. This is possible for example in anisotropic crystals or quantum Hall models accompanied by an in-plane magnetic field~\cite{Ian2024}.
This problem has been noted by previous authors~\cite{offertaler2019viscoelastic,rao2020hall} who dubbed it the \textit{Hall elastic modulus}. 

 {We note that one way to eliminate the Hall elastic modulus is to manually symmetrize the two Lie derivatives appearing in Eq.~\eqref{eq:D2}. This yields the definition of the second partial derivative in equation \eqref{eq:partial2Hpartiallambda}.
\begin{align}
    \frac{\partial^2 \hat H}{\partial \lambda_a \partial \lambda_b} &= \frac{1}{2}\left( {\mathcal L_{\overline v_a}}{\mathcal L_{\overline v_b}} + \mathcal L_{\overline v_b}\mathcal L_{\overline v_a}  \right)\hat H\\
    &= \frac{1}{2}( [\hat J^i_j,[\hat J^k_l,\hat H]]+[\hat J^k_l,[\hat J^i_j,\hat H]] ).
\end{align}} 
As we will argue below, however, this is insufficient to obtain the physically-relevant elastic response. 
We will now show how to calculate the physical contact term $\delta_\Lambda T$ and compare it to Eq.~\eqref{eq:Jacobi}.

\subsection{Present Formulation of the Elastic Contact Term}\label{sec:delta g T}

{In the previous subsection, we established that $(\Delta_\lambda \hat T_0)^{i~k}_{~j~l} \neq -\partial^2\hat H_\Lambda/\partial \lambda_i^j\partial \lambda^l_k$. Returning to equation \eqref{eq:deltaT}, we can now directly calculate $ (\delta_\Lambda\hat T)^{i~k}_{~j~l} \equiv \partial \hat T_{j}^i/\partial \Lambda_k^l = -\partial^2 \hat H_\Lambda/\partial \Lambda_i^j \partial\Lambda_k^l$ in terms of the Ward identity-derived stress tensor operator $(\hat{T}_0)^i_j$. A short calculation yields (for full details, see Appendix~\ref{sec:A0}, where we also verify the hyperelastic symmetry of $(\delta_\Lambda\hat T)_{~j~l}^{i~k} $)}
\begin{align}\label{eq:deltaTresult}
     (\delta_\Lambda\hat T)_{~j~l}^{i~k} &= (\Delta_\lambda \hat T_0)^{i~k}_{~j~l}- \delta^i_l (\hat T_0)^k_j\nonumber\\
\begin{split}
  &=- i[\hat{J}^k_l,(\hat T_0)^i_j] - \delta^i_l (\hat T_0)^k_j.
   \end{split}
\end{align}

\textcolor{black}{
It is worth noting that $\delta_\Lambda \hat T$ is equivalent to a directional derivative in the strain space parameterized by $\delta \Lambda$,
\begin{align}\begin{split}\label{eq:AbelianD}
   \nabla_{ \delta \Lambda_i^j} \mathcal  O_ \Lambda= \frac{\partial \mathcal O}{\partial \Lambda_i^j}.
    \end{split}
\end{align}
The space of strains parametrized by $\delta\Lambda$ forms an abelian group manifold $\simeq \mathbb R^{d^2}(+)$, equivalent to point-wise addition of matrices  $\delta\Lambda$ rather than matrix multiplication as was the case in $GL(d,\mathbb R)$ (See Figure \ref{fig:2}).}
Importantly, classical elasticity theory makes use of this abelian group structure by construction, {which is why taking multiple partial derivatives with respect to strain parameters does not lead to the complications regarding strain order as detailed in section \ref{sec:contact term comparison}.}

{From the perspective of transport on the non-abelian $GL(d,\mathbb R)$ manifold, equation \eqref{eq:deltaTresult} is equivalent to
\begin{align}
    (\delta_\Lambda \hat T_0)^{i~k}_{~j~l} &= \mathcal L_{\overline v^k_l} (\hat T_0)^i_j - \Gamma^{\binom{n}{m}}_{\binom{j}{i}\binom{l}{k}}( \hat T_0)^n_m.
\end{align}
With $\Gamma$ given by equation \eqref{eq:Gamma}.}

{Eq.~\eqref{eq:deltaTresult} gives us, via Eq.~\eqref{eq:Kubo}, the correct form of the contact term necessary in the Kubo formula for viscoelastic response. We see that, compared with the results of Refs.~\cite{bradlyn2012kubo,offertaler2019viscoelastic,rao2020hall}, there is an extra contribution to the contact term to convert $\Delta_\lambda \hat{T}_0$ to $\delta_\Lambda \hat{T}_0$. 
Substituting Eq.~\eqref{eq:deltaTresult} into Eq.~\eqref{eq:omegais0} allows for the expression of the first elasticity $A^{i~k}_{~j~l}$ in terms of the stress operator $\hat{T}_0$ and the strain generators $\hat{J}$, with the hyperelastic symmetry of the first elasticity made manifest. Furthermore, b}y combining our results in equation \eqref{eq:deltaTresult} with the extra factor implied by equation \eqref{eq:secondMod}, we can convert the first elasticity to the second elasticity {for isotropic systems (or more generally, systems that couple to strain only through the metric)}. As a reminder, all calculations of elastic moduli (bulk modulus, shear modulus, etc.) conventionally refer to components of the second elasticity.
The general formula in elastic equilibrium is 
\begin{align}
    (\delta_g \hat T)^{i~k}_{~j~l} &= \frac{1}{2}(\delta_\Lambda \hat T)^{i~k}_{~j~l} - \frac{1}{2}(\hat T_0)^{ik}g_{jl},\\
\label{eq:deltagT}
\begin{split}
    &= - \frac{i}{2}[\hat{J}^k_l,(\hat T_0)^i_j] - \frac{1}{2}\delta^i_l (\hat T_0)^k_j - \frac{1}{2}(\hat T_0)^{ik}g_{jl}.
   \end{split}
\end{align}
Where $ (\delta_g \hat T)^{ijkl}\equiv \partial \hat T^{ij}/\partial g_{kl}$.
Equation \eqref{eq:deltagT} manifestly possesses the symmetry $i\leftrightarrow j$ characteristic of the second elasticity [Eq. \eqref{eq:secondElasticity}], and can be shown to also possess symmetries under ($k\leftrightarrow l, ~ij\leftrightarrow kl$) by using the Ward identity [Eq. \eqref{eq:Ward}] and the Jacobi identity [Eq. \eqref{eq:Jacobi}].
{The relationship between $(\delta_g\hat T)^{ijkl}$ and the elastic modulus is given through the Kubo formula as
\begin{align}
    \begin{split}
        K^{ijkl} =- \langle \delta_g\hat T^{ijkl}\rangle_0- \frac{1}{2}\int dt' \chi_{TT}^{ijkl}( t'),
        \end{split}
\end{align}
which is determined from the first elasticity \eqref{eq:omegais0} and the relationship between the induced metric and the strain parameter.}
{Note that the result presented in equation \eqref{eq:deltagT} reduces to the formulation of the elastic modulus presented in \cite{bradlyn2012kubo} if and only if $\langle(\hat T_0)^i_j\rangle_0 = P \delta ^i_j.$}

\subsection{Example - Non-interacting Electron Fluids}

\label{sec:3b}

\begin{figure}
    \centering
    \includegraphics[width=1\linewidth]{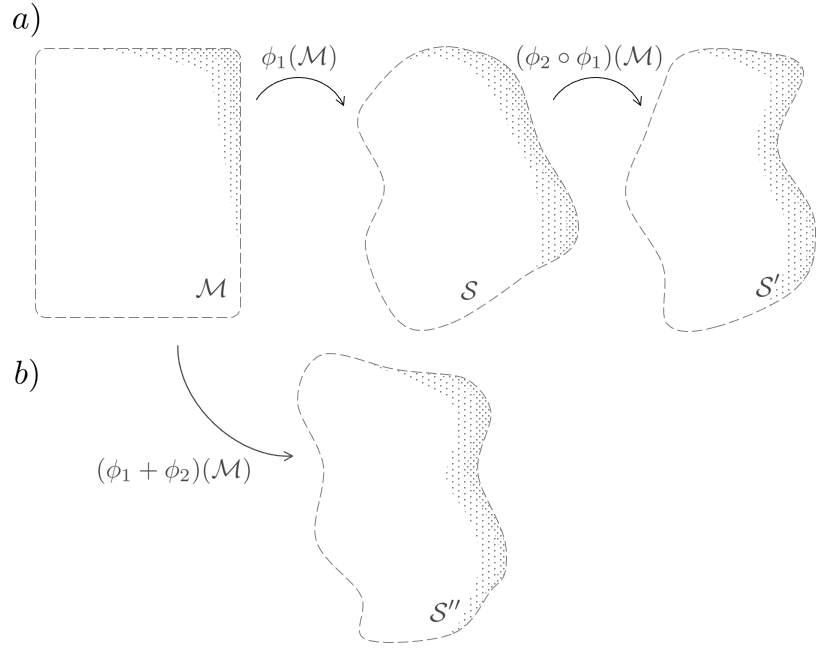}
    \caption{a) The natural method of combining strains acting on a quantum mechanical system is by composition, {which gives a representation of the} $GL(d,\mathbb R)$ group structure.  Multiple $\phi$ maps act one after the other where the lab manifold becomes the new reference manifold. b) The group structure of classical elasticity is abelian by relying on information about the reference manifold, i.e. $\phi$ acts on the constant $\mathcal M$. Because of this, successive strain maps $\phi_1$ and $\phi_2$ are both maps from $\mathcal{M}$ to $\mathcal{S}$ and so cannot be composed. However, exploiting the additivity of vectors, the point-wise addition $\phi_1+\phi_2$ is {(for small strains)} a well-defined strain.  { For arbitrarily large elements of the abelian strain space it is possible to create degenerate strains, i.e. a strain which collapses at least one dimension of the system. This pathology is a consequence of working in an abelian space (note this problem does not occur in the $GL(d,\mathbb R)$ strain group) and is mostly ignored in the literature by considering ``small strains."}
   } 
    \label{fig:2}
\end{figure}

For a non-interacting quadratic Hamiltonian the commutator of $\hat{J}$ and $\hat T_0$ is given by
\begin{align}\label{eq:JTcom}
    -i [\hat J^{kl},\hat T_0^{ij}] = - g^{ki}\hat T_0^{lj} - g^{kj}\hat T_0^{il}.
\end{align}
{Using this in Eq.~\eqref{eq:deltagT}, we find that the contact term entering the definition of the second elasticity is}
\begin{align}\label{eq:Symmetric}
   (\delta_g \hat T)^{ijkl}&= -\frac{1}{2}\bigg( g^{ik}\hat T_0^{lj} + g^{kj}\hat T_0^{il} + g^{il} \hat T_0^{kj} + g^{jl}\hat T_0^{ik} \bigg).
\end{align}
In Appendix \ref{sec:Review} we further derive the contact term for a general rotationally-invariant interacting Hamiltonian.
When combined with the $\omega = 0$ component of the integral term, Eq.~\eqref{eq:JTcom} gives $K^{ijkl}$ for non-interacting electrons. 

For isotropic systems, the Kubo formula of equation \eqref{eq:Kubo} can be re-expressed in terms of variations of the metric as
\begin{align}
    \langle \hat T^{ij}(t)\rangle - \langle \hat T_0^{ik}\rangle_0 &= \langle \delta_g\hat T^{ijkl}\rangle_0 \delta g_{kl}(t)+ \frac{1}{2}\int dt' \chi_{TT}^{ijkl}\delta g_{kl}(t'), \nonumber\\
    &=- K^{ijkl} \delta g_{kl} - \eta^{ijkl}\partial _t\delta g_{kl} + ...
\end{align}

As a quick check, we calculate the second derivative of a non-interacting Hamiltonian with respect to a spatially independent metric and imagine turning on a strong magnetic field to open a gap so that strains can be applied adiabatically. 
{As we mentioned in section \ref{sec:review}, there are additional subtleties for systems in a nonzero external magnetic field. This can be seen by the gauge variance of equation \eqref{eq:StrainGen}. Ref.~\cite{bradlyn2012kubo} generalizes the strain generator to systems with non-zero vector potential, and the results of our present work may be extrapolated by using their version of the strain generator in place of the one defined in equation \eqref{eq:StrainGen}.}
We use adiabatic response theory to deduce~\cite{avron1987adiabatic,read2011hall} that
\begin{align}
    \langle \hat H(g) \rangle &= \sum_{n\in \text{occ}} \int \,d x\, \sqrt{g}~\psi^*_n(x) \frac{g^{ij}}{2m}\hat \pi_i\hat \pi_j \psi_n(x), \\
     K^{ijkl}
   &= \frac{1}{2}(g^{ik}T^{jl}+ g^{il }T^{jk} + g^{jk}T^{il}+ g^{jl}T^{ik})\nonumber \\
   &-\sum_{n \in \text{occ}} \int d^2 x  \bigg( \frac{\partial \Psi^*_n(x)}{\partial g_{kl}}~\hat T_0^{ij} ~  \Psi_n(x) + h.c.\bigg),\label{eq:K_non_int}
\end{align}
where $\psi_n(x) = \langle x | \psi_n\rangle$ is an eigenstate of the Hamiltonian, $\Psi_n(x)= (\det g)^{1/4}~\psi_n(x)$, the sum is over occupied states, the stress tensor of the non-interacting particles in magnetic field is $(\hat T_0)^{ij} = \frac{g^{ik}g^{jl}}{2m}\{\hat\pi_k,\hat\pi_l\}$, and $\hat \pi$ is the gauge-invariant kinetic momentum. Here, we used the Hellman-Feynman theorem along with the identity $\partial g^{ij}/\partial g_{kl} = -\frac{1}{2}(g^{ik}g^{jl} + g^{jk}g^{il})$.
The second term {in Eq.~\eqref{eq:K_non_int}} corresponds to the integral term in the Kubo formula while the first is exactly what we calculate {for the contact term} in equation \eqref{eq:Symmetric}.
By computing the $\omega = 0$ component of the Kubo formula for the free, quadratic Hamiltonian filled to $\nu$ Landau levels we {recover the known result} $K^{ijkl} =  E_0  g^{ij}g^{kl}$ where $E_0 = \frac{N_0 \nu B}{ m}(\nu + 1/2)$ is the total energy of the ground state with  level degeneracy $N_0$. 
This result is consistent with the interpretation of the quantum Hall system as a fluid since the elastic modulus $K$ has vanishing shear modulus.
The manifest symmetry of our framework ensures there are no nonzero Hall elastic moduli, since the hyper-elastic symmetry of Eq.~\eqref{eq:Symmetric} is automatically enforced.

Crucially, the linear Hall viscosity is proportional to a Berry curvature in the space of strains~\cite{avron1995viscosity} and therefore originates entirely from the integral term in the Kubo formula. 
Thus, the results of prior works which utilize the viscoelastic Kubo formula regarding the Hall viscosity are not updated by any corrections made in this work. 
However, by extending the Kubo formula to second order in strain fields, one can in principle calculate the non-linear viscosity which may depend on the diamagnetic term\,\cite{jain2025accoustic,jain2025nonlinearoddviscoelasticeffect}.

\section{Stress-Strain Form of the Kubo Formula and Sum rules}
\label{sec:SumRule}
We have established in Eqs.~\eqref{eq:Kubo} and \eqref{eq:deltaTresult} how to modify the contact term in the Kubo formula for stress response in order to make contact with classical elasticity. We now turn to a direct comparison with the Kubo formula for viscosity derived in Refs.~\cite{bradlyn2012kubo,Randeria_Viscosity}. In those references, linear response of the stress to the strain rate $\dot\Lambda^{i}_j$ is considered, and several forms of the Kubo formula are derived by making use of the Ward identity [Eq.~\eqref{eq:Ward}]. We will show that our main results imply that there is a new contact term in the Kubo formula when expressed in terms of the stress tensor and the strain generators. The new contact term will prove to be significant for the viscosity sum rules which in principle can be experimentally probed.

{For completeness, we will carefully derive the Kubo formula for the linear response of the stress to strain rate. To be consistent with the terminology of Ref.~\cite{bradlyn2012kubo}, we refer to this as the ``stress-strain'' form of the Kubo formula. We start with the perturbed Hamiltonian from Eq.~\eqref{eq:perturbedHam}, which we can rewrite using the Ward identity as
}
\begin{align}\label{eq:secondOrderH}
\begin{split}
   \hat  H_\Lambda = \hat H_0 - \delta\Lambda_{i}^j \bigg[i &[\hat{J}^i_j, \hat H_0 ] -i\delta\Lambda_k^l [\hat{J}^k_l,(\hat T_0)^i_j]\bigg]\\
   &+  \delta^i_l (\hat T_0)^k_j\delta\Lambda^j_i\delta\Lambda^l_k + \mathcal O(\delta \Lambda^3)
   \end{split}
\end{align}
The last term is the new result of this work as it does not appear in Refs. \cite{bradlyn2012kubo} or \cite{offertaler2019viscoelastic} {for reasons discussed in Sec.~\ref{sec:contact term comparison}}. The discrepancy arises because in those works, the authors expand the Hamiltonian in terms of $\lambda$ instead of $\delta\Lambda$.

In the interaction picture, the time evolution of the ground state density matrix $\rho$ is governed by the perturbation $H'$ (see equation \eqref{eq:Hprime}),
\begin{align}
    \hat \rho(t) =\hat  U(t) \hat \rho_0 \hat U^\dag (t), \\
    \hat U(t) = T\exp\left(-i \int _{t_0}^t \hat H' (t') dt'\right),
\end{align}
where $T$ denotes to a time ordering. To first order in $\delta \Lambda$ the interacting evolution operator is 
\begin{align}
    \hat U(t) = \mathbb 1 -i \int _{t_0}^t \delta \Lambda_i^j \frac{d \hat{J}^i_j}{dt }.
\end{align}
Integrating by parts yields 
\begin{align}\label{eq:interactionUnitary}
    \hat U(t) = \mathbb 1 +i \int _{-\infty}^t dt'  \hat{J}^i_j\frac{d \Lambda_i^j }{dt' } - i \delta\Lambda_i^j(t)  \hat{J}^i_j\,,
\end{align}
{where we took the limit $t_0\rightarrow-\infty$. }

The stress-strain Kubo formula can then be derived where the last term in equation \eqref{eq:interactionUnitary} cancels the commutator in the second order expansion of equation \eqref{eq:secondOrderH}.
Computing the average of the stress tensor, we find
\begin{align}\label{eq:StressStrainKubo}
    \langle \hat T^i_j (t) \rangle - \langle (\hat T_0)^i_j \rangle_0=&-\delta\Lambda_k^i(t) \langle(\hat T_0)_j^k\rangle_0\\
      &
     - \int _{- \infty}^\infty dt'    (\chi^{TJ})^{i~k}_{~j~l}(t-t')\frac{d \Lambda^l_k}{dt'}\,, \nonumber
\end{align}
where
\begin{align} 
    & (\chi^{TJ})^{i~k}_{~j~l}(t)  =- i \Theta(t) \langle [(\hat T_0(t)  )^i_j, \hat{J}^k_l (0)] \rangle_0\,.
\end{align}

Interestingly, equation \eqref{eq:StressStrainKubo} implies that to linear order in $\delta\Lambda$ the elastic response is given by the stress-strain integral and an expansion of the equilibrium stress contracted with the inverse strain operator.
\begin{align}
    \langle \hat T^i_j (t) \rangle 
     \approx\, &  \,(\Lambda^{-1}(t))_k^i \langle(\hat T_0)^k_j \rangle_0  \\
     &- \int _{- \infty}^\infty dt'    (\chi^{TJ})^{i~k}_{~j~l}(t-t')\frac{d \Lambda^l_k}{dt'} + \mathcal O(\delta\Lambda^2).\nonumber
\end{align}
{Finally, we can take the Fourier transform to frequency space to arrive at}
\begin{align}\label{eq:StressStrainKubo1}
    \langle \hat T^i_j (\omega) \rangle - (\Lambda^{-1}(\omega))_k^i \langle(\hat T_0)^k_j \rangle_0= -( \chi^{TJ})^{i~k}_{~j~l}(\omega ) (  i \omega^+
   ) \delta\Lambda_k^l(\omega) \,
\end{align}
where
\begin{align}
    & ( \chi^{TJ})^{i~k}_{~j~l}(\omega )  =- i \lim_{\epsilon\rightarrow 0^+}\int_0^\infty dte^{i \omega^+ t} \langle [(\hat T_0(t)  )^i_j, \hat{J}^k_l (0)] \rangle_0\,.
\end{align}
Here we have defined $\omega^+ = \omega + i \epsilon$ with $\epsilon$ {being} a positive infinitesimal to enforce causality. Compared with the stress-strain Kubo formulas of Refs.~\cite{bradlyn2012kubo,offertaler2019viscoelastic}, we highlight the additional contact term on the left hand side of Eq.~\eqref{eq:StressStrainKubo1}.

Using causality, we can now define a spectral density representation of the response. First, recall that the stress-strain response function $( \chi^{TJ})^{i~k}_{~j~l}(\omega )$ can be written as
\begin{equation}\label{eq:integralpartofspecdensity}
    \chi^{TJ}( \omega^+)^{i~k}_{~j~l} = \int \frac{d\omega'}{\pi } \frac{\chi''(\omega')^{i~k}_{~j~l}}{\omega' - \omega^+},
\end{equation}
with
\begin{align}
    \chi''(\omega)^{i~k}_{~j~l} =  - \frac{i}{2}\int_{-\infty}^{\infty} dt  e^{ i \omega t}\langle [(\hat T_0(t)  )^i_j, \hat{J}^k_l (0)] \rangle_0
\end{align}
the antihermitian part of $\chi^{TJ}( \omega^+)^{i~k}_{~j~l}$~\cite{forster1975hydrodynamic}.

We can combine Eq.~\eqref{eq:integralpartofspecdensity} with the Fourier representation of the contact term in Eq.~\eqref{eq:StressStrainKubo1} to obtain the full spectral function for the stress-strain response. Inserting a Dirac delta function, we define
\begin{align}
       \langle \hat T^i_j (\omega) \rangle
       &=  - (i \omega^+)\delta \Lambda^l_k(\omega) \int \frac{d\omega'}{\pi } \bigg( \frac{\widetilde \chi'' ( \omega')^{i~k}_{~j~l}}{\omega' - \omega^+}\bigg)\,,
\end{align}
with
\begin{align}
       \widetilde \chi'' ( \omega')^{i~k}_{~j~l} &= \chi''(\omega')^{i~k}_{~j~l}+ \pi \delta(\omega')\delta^i_l \langle(\hat T_0)_j^k\rangle_0.\label{eq:fullspectraldensity}
\end{align}

The spectral representation in Eq.~\eqref{eq:fullspectraldensity} allows for the derivation of families of sum rules for the stress-strain response. The simplest sum rule comes from integrating $\widetilde \chi'' ( \omega')^{i~k}_{~j~l}$ over all frequency. With the help of Eqs.~\eqref{eq:integralpartofspecdensity} and Eq.~\eqref{eq:deltaTresult} we find
\begin{align}
     \int_{-\infty}^{\infty} \frac{d\omega}{\pi } (i\omega)^0\widetilde \chi''(\omega)^{i~k}_{~j~l} = - (\langle \delta _\Lambda \hat T\rangle_0)^{i~k}_{~j~l},
\end{align}
which manifestly possesses the hyper-elastic symmetry. {This shows that the} contact term is potentially experimentally measurable by a dynamical measurement of the frequency dependent viscosity.
This is a similar measurement to that of frequency dependent acoustic waves~\cite{Kazemirad2013,LORENZ2014565}. 

\section{Discussion}
\label{sec:discussion}

{In this work, we critically re-examined the quantum formulation of elasticity with a particular eye to subtleties that arise in anisotropic fluids and solids. We highlighted an unexpected subtlety in defining contact terms in the stress tensor of quantum systems that manifests only when the ground state of the system breaks spatial isotropy. We showed that in these cases, a naive application of the Kubo formula for viscosity from Refs.~\cite{bradlyn2012kubo,offertaler2019viscoelastic} yields a spurious odd elastic response whose origin is fundamentally geometric. Rather than corresponding to the physical response of the system to strains, it reflects the fact that taking the second variation of the Hamiltonian with respect to strain is nontrivial since the group of strains is nonabelian. We argue by analogy with classical elasticity that this geometric contribution to the Kubo formula, while interesting, does not correspond to the physically measurable elastic modulus of the fluid. Through a careful treatment of strain perturbations we show how to modify the contact terms in the Kubo formula to produce the physical elastic moduli.}

{Our work has several implications for the study of elasticity in quantum fluids. First, our main result that the physical elastic modulus satisfies the hyper-elastic relation shows that, as argued classically in Ref.~\cite{OddElasticity2024}, the physical odd elastic modulus must vanish in energy-conserving quantum fluids. Applied to the case of a quantum Hall fluid in a tilted field, this establishes that the ``Hall elastic modulus'' of Ref.~\cite{offertaler2019viscoelastic} does not correspond to any elastic work done on the system, but rather reflects the interplay between the Lie group geometry of strains and the anisotropic stress in the ground state. More generally, we show that the Kubo formula for the physical elastic modulus implies a new viscosity sum rule, which (in contrast to Ref.~\cite{bradlyn2012kubo}) explicitly respects hyperelastic symmetry even for anisotropic fluids and solids. This sum rule is exact, and holds even for strongly-interacting systems. We thus expect it could be experimentally investigated in, e.g. ultracold gasses in a magnetic field.}

{As interest turns towards the role of topological aspects of non-linear response functions, our results becomes a necessary guide for the complete construction of a non-linear (or second order) viscoelastic response theory. Such a theory describes the effect of a system given two strain perturbations. For example, the second order viscosity becomes the sum of two terms~\cite{SecondOrderSumRules,jain2025accoustic,jain2025nonlinearoddviscoelasticeffect}:  a Kubo integral over two time parameters of nested commutators with the strain generator and the stress tensor $[\hat T^i_j(0),[\hat T^k_l(t_1),\hat T^n_m(t_1+t_2)]]$, and a Kubo integral over one time parameter of the commutator of the diamagnetic term and the stress tensor $[ (\delta_\Lambda \hat T)^{i~k}_{~j~l}(0), \hat T^n_m(t)]$. The careful treatment of $(\delta_\Lambda \hat T)^{i~k}_{~j~l}$ that we outline here is necessary to obtain physically meaningful results.}

{Beyond elasticity, our work has important implications for analyzing contact terms in linear and nonlinear response. To see this, we can consider the general case of a Hamiltonian with a non-abelian set of symmetry generators. The generalization of Eq.~\eqref{eq:Ward} implies that the conserved current in zero external field is given by the commutator of the Hamiltonian with the symmetry generators. To next order in the external field, we know that there are generally ``diamagnetic'' contributions to the current, which are responsible for contact terms in the Kubo formula. Following our discussion in Sec.~\ref{sec:contact term comparison}, in general the diamagnetic currents are \emph{not} simply double commutators of the Hamiltonian with respect to the symmetry generators.
}

{For the ordinary $U(1)$ number current, this complication does not arise; the symmetry group is abelian, Fourier components of the charge density operator all commute, and hence the diamagnetic charge current can be simply expressed as the double commutator of the Hamiltonian with the charge density operator~\cite{watanabe2020generalized,mckay2024charge,bradlyn2024spectral}.
The situation is more complicated if one were to consider charge transport in a low-energy projected model, such as is relevant for topology in moir\'e systems and fractional Chern insulators. This is because, while the density operators commute, the density operators projected to a set of low-energy bands satisfy a more complicated non-abelian algebra. Our work shows that the projected linear and nonlinear response of the charge current must be carefully treated. In particular, our Lie derivative approach to contact terms provides a necessary perspective for generalizing the projected sum rules of Refs.~\cite{Mendez_Valderrama_2024,mao2025lowenergy} to off-diagonal components of the conductivity tensor and quantum metric.}

\section*{Acknowledgments}
The authors thank P. Nair for bringing the connection to Refs.~\cite{dashen1969chiral, WeinbergVol2} to our attention. I. O. and B. B. acknowledge the support of the National Science Foundation under grant No. DMR-2510219.

\appendix
\onecolumngrid

\section{Review of Quantum Hydrodynamics and Elasticity}
\label{sec:Review}

{In this appendix, we review the first-quantized  formulation of stress and strain. Our logic largely mirrors that of} Ref. \cite{bradlyn2012kubo}.
The stress tensor density $\tau^i_{~j}$ is the {intensive viscoelastic observable} whose gradient sets the bulk force density of an elastic or hydrodynamic material~\cite{landau_elasticity}.
We distinguish between the local force density generated by the stress $f^{\text{int}}_j=- \partial_i \tau^i_{~j}$ and the force density generated by external fields $f^{\text{ext}}_i$. 
The continuity equation expressing the conservation of momentum is
\begin{align}\label{eq:MomCont2}
    \partial_t g_i + \partial_j  \tau^j_{~i} =  f^\text{ext}_i,
\end{align}
where $g_i$ is the momentum density.

Let the momentum density be the expectation of the total momentum operator on an $N$ particle Hilbert space,
\begin{align}\label{eq:app_momentum_density}
    g_i(\textbf{r}) = \langle \frac{1}{2}\sum_n \{\delta(\hat {\textbf{x}}_n - \textbf{r}) , \hat \pi^n_i\}  \rangle,
\end{align}
{where $\hat{\pi}^n_i$ is the $i$-th component (kinetic) momentum operator for particle $n$, and $\hat{\mathbf{x}}_n$ is the position operator for particle $n$. Similarly, the number density operator is}
\begin{align}
    \rho(\textbf{r}) = \sum_n \langle \delta(\hat {\textbf{x}}_n - \textbf{r})\rangle,
\end{align}
and the external force density is
\begin{align}
    f^{\text{ext}}_i(\textbf{r}) = \frac{1}{2}\sum_n \{F^{\text{ext}}_i (\hat{\textbf{x}}_n, \hat{\mathbf{\pi}}^n), \delta (\hat{\textbf{x}}_n - \textbf{r})\}.
\end{align}

{We can take the Fourier transform of Eq.~\eqref{eq:MomCont2} and expand order-by-order in wavevector to obtain expressions for the forces and stresses on the system. Using Eq.~\eqref{eq:app_momentum_density}, at} zeroth order in {wavevector we recover Ehrenfest's theorem.}
\begin{align}\label{eq:Ehrenfest}
    \partial_t  \sum_n \langle \hat \pi^n_i \rangle =  \sum_n \langle F^{\text{ext}}_i(\hat{\textbf{x}}_n,\hat{\mathbf{\pi}}^n)\rangle
\end{align}
{The rate of change of total momentum is given by the total external force acting on the system. We note that in the presence of an external magnetic field, the Lorentz force acting on the particles is included in $\bf{F}^\mathrm{ext}$.}
Note {also} that the internal forces disappear in the sum over all particles by Newton's third law.
{Moving next to first order in $q$, we find that}
\begin{align}\label{eq:MomContq}\begin{split}
       - \frac{1}{2} \partial _t  \sum_n \langle \{\hat  x^j_n  ,\hat  \pi_i^n\}  \rangle  + \tau_{ji}(\textbf{q } = 0) = - \frac{1}{2} \sum_n \langle \{\hat x_n^j ,F_i^{\text{ext}}(\hat{\textbf{x}}_n, \hat{\mathbf{\pi}}^n)\}\rangle.\end{split}
\end{align}

Take note that the first term is the time derivative of the (zero magnetic field) strain generator presented in equation \eqref{eq:StrainGen}. 
The combination of equation \eqref{eq:MomContq} with the definition of the integrated stress tensor $T _{ij}= \tau_{ij}(\textbf{q}=0) = \int \sqrt{g}\,d^d \textbf{r}~ \tau_{ij}(\textbf{r})$ gives the following as a definition of the extensive stress:  
\begin{align}\begin{split}\label{eq:extensivestress1}
    T_{ij} = \frac{1}{2} \sum_{n} \langle \{\partial_t \hat x^i_n,\hat \pi_j^n\}\rangle + \frac{1}{2}\langle\{\hat x^i_n,\partial_t \hat \pi^n_j\} \rangle - \frac{1}{2}\langle \{ \hat x^i_n, F^{\text{ext}}_j(\hat{\textbf{x}}_n, \hat{\mathbf{\pi}}^n)\}\rangle.
    \end{split}
\end{align}

{For Hamiltonians with momentum-independent interactions and external potentials, the first term in Eq.~\eqref{eq:extensivestress1} includes contributions} exclusively from the kinetic energy, because all other terms in the Hamiltonian commute with the position operator. The second line may be simplified with equation \eqref{eq:Ehrenfest} by defining 
\begin{align}
     \partial_t \hat \pi_i^n =  F^{\text{ext}}_i(\hat{\textbf{x}}_n, \hat{\mathbf{\pi}}^n) + F^{\text{int}}_i(\hat{\textbf{x}}_n, \hat{\mathbf{\pi}}^n).
\end{align}
We then have that
\begin{align}\label{eq:extensivestress2}
    T^i_{~j}= \frac{1}{2} \sum_{n} \langle \{\partial_t \hat x^i_n,\hat \pi_j^n\}\rangle + \frac{1}{2}\langle \{\hat x_n^i , F^{\text{int}}_j(\hat{\textbf{x}}_n, \hat{\mathbf{\pi}}^n)\}\rangle.
\end{align}
{Let us now specialize to the case where the Hamiltonian is isotropic (in the absence of external electromagnetic fields), i.e. depends only on the geometry of space through the dependence on a single background metric that we will take to be Euclidean at the end of the calculation. The most general such nonrelativistic } Hamiltonian {with pair-wise interactions} coupled to an external electric and magnetic field is given by 
\begin{align}
    \hat H = \sum_n\frac{\hat \pi^n_i g^{ij}\hat \pi^n_j}{2m} +\frac{1}{2} \sum_{n \neq m} V(|\hat {\textbf{x}}_m - \hat {\textbf{x}}_n|) - E_i \sum_n \hat x_n^i.
\end{align}
We find that the extensive stress is given by
\begin{align}
    T_{j}^i = \frac{1}{2m}\sum_n \langle \{g^{ik}\hat \pi^n_k,\hat \pi^n_j\}\rangle- \frac{1}{2}\sum_{n \neq m} \langle \{ \hat x_n^i, \frac{\partial }{\partial \hat  x_n^j } V(|\hat {\textbf{x}}_n - \hat {\textbf{x}}_m|)\}\rangle .
\end{align}

{We have extensively discussed the kinetic contribution to the stress in Sec.~\ref{sec:3b}. It is illuminating now to} take specific care to simplify the expression coming from the inter-particle interactions. {A similar} derivation is carried out in Refs. \cite{bradlyn2012kubo,irving1950statistical,cooper1997thermoelectric}.
First, note that for arbitrary metrics $|\hat {\textbf{x}}_n - \hat {\textbf{x}}_m|$ represents the geodesic distance between the two particles, and, for the flat metrics considered here, the geodesic is a straight line. 
{The interaction contribution to the stress can be shown to depend only on the separation between particles through a relabeling of indices, yielding}
\begin{align}
    \frac{1}{2}\sum_{n \neq m} \langle \{\hat  x_n^i, \frac{\partial }{\partial \hat x_n^j } V(|\hat {\textbf{x}}_n - \hat {\textbf{x}}_m|)\}\rangle  &=  \frac{1}{4}\sum_{n \neq m} \langle \{ (\hat x_n^i -\hat  x_m^i), \frac{\partial }{\partial \hat x_n^j } V(|\hat {\textbf{x}}_n- \hat {\textbf{x}}_m|)\}\rangle \\ &= \frac{1}{2}\sum_{n \neq m} \left\langle (\hat x_n^i -\hat  x_m^i) V'(|\hat {\textbf{x}}_n- \hat {\textbf{x}}_m|)\frac{g_{jl}(\hat x_n^l- \hat x_m^l)}{|\hat {\textbf{x}}_n - \hat {\textbf{x}}_m|}\right\rangle ,\label{eq:intstress}
\end{align}
where we have used 
\begin{align}
|\hat {\textbf{x}}_n - \hat {\textbf{x}}_m|= \sqrt{g_{ij}(\hat x_n^i-\hat x_m^i)(\hat x_n^j - \hat x_m^j)}.
\end{align}

Thus, the full expression for the extensive stress is
\begin{align}\label{eq:extensivestress3}
\begin{split}
     T^i_{~j}= \frac{1}{2m} \sum_{n} \langle \{g^{ik}\hat \pi_k^n,\hat \pi_j^n\}\rangle - \frac{1}{2}\sum_{n \neq m}\left\langle V'(|\hat {\textbf{x}}_n - \hat {\textbf{x}}_m|)\frac{(\hat x^i_n - \hat  x^i_m)g_{jl}(\hat x^l_n - \hat x^l_m)}{|\hat {\textbf{x}}_n - \hat {\textbf{x}}_m|}\right\rangle .
     \end{split}
\end{align}

In elastic equilibrium $g = \delta$ and in the absence of external forces, equation \eqref{eq:extensivestress3} is equivalent to the Ward identity of equation \eqref{eq:Ward}.
For this reason we say the strain generator $\hat J$ generates the stress through a time derivative.

{Using the results of Sec.~\ref{sec:delta g T}, we now calculate the interaction contribution to the elastic contact term $\delta_g\hat T $} for the system whose stress is given by equation \eqref{eq:extensivestress3}. 
First, let us decompose the stress into a part coming from the kinetic energy $\hat T_K$ and a part coming from the particle-particle potential $\hat T_V.$
The elastic modulus of the kinetic stress term is given by equation \eqref{eq:Symmetric} in the main text {(as the example is that of a non-interacting gas of particles)}.
Therefore, it is only necessary to calculate the elastic term of the interaction stress. The commutator of $\hat T_V $ and $\hat J$ is 
\begin{align}
  [\hat{J}^k_l,( \hat T_V)^i_j] &=  \frac{1}{4}\sum_r\sum_{n \neq m}\bigg[   \{\hat  x^k_r , \hat \pi_l^r\}, V'(|\hat {\textbf{x}}_n - \hat {\textbf{x}}_m|)\frac{(\hat x^i_n - \hat x^i_m)g_{jq}(\hat x^q_n - \hat x^q_m)}{|\hat {\textbf{x}}_n - \hat {\textbf{x}}_m|} \bigg]\\
  &=  \frac{1}{2}\sum_r\sum_{n \neq m}x_r^k\bigg[    \pi_l^r, V'(|\hat {\textbf{x}}_n - \hat {\textbf{x}}_m|)\frac{(\hat x^i_n - \hat x^i_m)g_{jq}(\hat x^q_n - \hat x^q_m)}{|\hat {\textbf{x}}_n - \hat {\textbf{x}}_m|} \bigg]\\
  \begin{split}
   &= - \frac{i}{2}\sum_{n \neq m} \bigg( V''(|\hat {\textbf{x}}_n - \hat {\textbf{x}}_m|) - \frac{V'(|\hat {\textbf{x}}_n - \hat {\textbf{x}}_m|)}{|\hat {\textbf{x}}_n -\hat {\textbf{x}}_m| }\bigg) \frac{(\hat x^i_n - \hat x^i_m)g_{jq}(\hat x^q_n - \hat x^q_m)}{|\hat {\textbf{x}}_n - \hat {\textbf{x}}_m|} \frac{(\hat x^k_n - \hat x^k_m)g_{wl}(\hat x^w_n - \hat x^w_m)}{|\hat {\textbf{x}}_n - \hat {\textbf{x}}_m|} \\
   &-   \frac{i}{2}\sum_{n \neq m} V'(|\hat {\textbf{x}}_n - \hat {\textbf{x}}_m|)\frac{(\hat x^k_n - \hat x^k_m)g_{jq}}{|\hat {\textbf{x}}_n - \hat {\textbf{x}}_m|} \left( \delta^i_l (\hat x^q_n -\hat  x^q_m ) + \delta^q_l (\hat x^i_n - \hat x^i_m)   \right).
   \end{split}
\end{align}

Now we may calculate $\delta _g \hat T_V$ using equation \eqref{eq:deltagT}.
\begin{align}
   &( \delta _g \hat T_V)^{ijkl} = \\
   \begin{split}
   &- \frac{1}{4}\sum_{n \neq m} \bigg( V''(|\hat {\textbf{x}}_n - \hat {\textbf{x}}_m|) - \frac{V'(|\hat {\textbf{x}}_n - \hat {\textbf{x}}_m|)}{|\hat {\textbf{x}}_n -\hat {\textbf{x}}_m| }\bigg) \frac{(\hat x^i_n - \hat x^i_m)(\hat x^j_n -\hat  x^j_m)}{|\hat {\textbf{x}}_n - \hat {\textbf{x}}_m|} \frac{(\hat x^k_n - \hat x^k_m)(\hat x^l_n -\hat  x^l_m)}{|\hat {\textbf{x}}_n - \hat {\textbf{x}}_m|} \\
   &-   \frac{1}{4}\sum_{n \neq m} V'(|\hat {\textbf{x}}_n - \hat {\textbf{x}}_m|)\bigg( \frac{(\hat x^k_n - \hat x^k_m)g^{il}(\hat x^j_n -\hat  x^j_m )+  (\hat x^l_n - \hat x^l_m ) g^{kj}(\hat x^i_n - \hat x^i_m)}{|\hat {\textbf{x}}_n - \hat {\textbf{x}}_m|}  \bigg)\\
    &-   \frac{1}{4}\sum_{n \neq m} V'(|\hat {\textbf{x}}_n - \hat {\textbf{x}}_m|)\frac{(\hat x^i_n - \hat x^i_m)g^{lj} (\hat x^k_n - \hat x^k_m)}{|\hat {\textbf{x}}_n - \hat {\textbf{x}}_m|}\\
     &+ \frac{1}{4}\sum_{n \neq m} V'(|\hat {\textbf{x}}_n - \hat {\textbf{x}}_m|)\frac{(\hat x^i_n - \hat x^i_m)g^{kj}(\hat x^l_n -\hat  x^l_m)}{|\hat {\textbf{x}}_n - \hat {\textbf{x}}_m|} + \frac{1}{4}\sum_{n \neq m} V'(|\hat {\textbf{x}}_n - \hat {\textbf{x}}_m|)\frac{(\hat x^k_n - \hat x^k_m)g^{il}(\hat x^j_n -\hat  x^j_m)}{|\hat {\textbf{x}}_n - \hat {\textbf{x}}_m|}\\
      &+ \frac{1}{4}\sum_{n \neq m} V'(|\hat {\textbf{x}}_n - \hat {\textbf{x}}_m|)\frac{(\hat x^i_n - \hat x^i_m)g^{lj}(\hat x^k_n - \hat x^k_m)}{|\hat {\textbf{x}}_n - \hat {\textbf{x}}_m|}
      \end{split}\\
      &= - \frac{1}{4}\sum_{n \neq m} \bigg( V''(|\hat {\textbf{x}}_n - \hat {\textbf{x}}_m|) - \frac{V'(|\hat {\textbf{x}}_n - \hat {\textbf{x}}_m|)}{|\hat {\textbf{x}}_n -\hat {\textbf{x}}_m| }\bigg) \frac{(\hat x^i_n - \hat x^i_m)(\hat x^j_n -\hat  x^j_m)}{|\hat {\textbf{x}}_n - \hat {\textbf{x}}_m|} \frac{(\hat x^k_n - \hat x^k_m)(\hat x^l_n - \hat  x^l_m)}{|\hat {\textbf{x}}_n - \hat {\textbf{x}}_m|} \label{eq:ElasticV}
\end{align}
Interestingly, most of the terms cancel and we are left with the relatively simple expression \eqref{eq:ElasticV}.
The contact term for the elasticity $\delta_g \hat T$ is the sum then of the expressions in \eqref{eq:Symmetric} and \eqref{eq:ElasticV}.
{Recall from Secs.~\ref{sec:review} and \ref{sec:Kubo} that the quantity above represents the second derivative of the Hamiltonian with respect to the metric.
\begin{align}
  ( \delta_g \hat T_V) ^{ijkl} = - \frac{1}{\sqrt{g}}\frac{\partial ^2\hat H_\Lambda}{\partial g_{ij}\partial g_{kl}}
\end{align}
A useful check on the result above is to note that if $V(\textbf{x})\propto x^i g_{ij}x^j$ then $\delta _g \hat T_V = 0$ for the simple reason that the second derivative of $V$ with respect to the metric is zero. Unsurprisingly, without including each of the counter-terms calculated in this paper and displayed in equation \eqref{eq:deltagT}, the result formula for $\delta_g \hat T_V$ takes on explicit metric dependence for the quadratic particle potential}.

{In the next subsection, we will discus the formulation of the first P-K stress and first elasticity in the presence of anisotropy in the form of a band mass tensor and electric permittivity tensor which are not proportional to the spatial metric. }

\subsection{Anisotropy}
\label{sec:anisotropy}
The identity expressed in equation \eqref{eq:ElasticV} uses the property that the  distance $|\hat {\textbf{x}}_n - \hat {\textbf{x}}_m|$ used in the potential is the true geodesic distance, but anisotropy can arise in the interaction term {through the appearance of a space-independent} anisotropic permittivity tensor. Likewise, an effective band mass may introduce anisotropy into the kinetic term.
For these non-Cauchy materials, the spatial metric is no longer the only relevant symmetric 2-tensor used in calculating distance or contracting indices. Therefore, the relationship between the strain and the spatial metric breaks down such that the second P-K stress can not be found by taking derivatives with respect to the spatial metric.

The first P-K stress, however,  is always definable as the first derivative with respect to $\Lambda$ evaluated at $\Lambda = \delta.$
For an arbitrary  {non-relativistic} Hamiltonian, {we find from Eq.~\eqref{eq:FirstPiola2} that} 
\begin{align}\label{eq:FirstAniso}
   {T_{j}^i}= \frac{1}{2} \sum_{n} \langle \{(\widetilde m^{-1})^{ik}\hat \pi_k^n,\hat \pi_j^n\}\rangle - \frac{1}{2}\sum_{n \neq m}\left\langle V'(|\hat {\textbf{x}}_n - \hat {\textbf{x}}_m|_\epsilon)\frac{(\hat x^i_n - \hat x^i_m)\epsilon_{jl}(\hat x^l_n - \hat x^l_m)}{|\hat {\textbf{x}}_n - \hat {\textbf{x}}_m|_\epsilon}\right\rangle,
\end{align}
where
\begin{align}
    |\hat {\textbf{x}}_n - \hat {\textbf{x}}_m|_\epsilon = \sqrt{\epsilon_{ij}(\hat x^i_n - \hat x^i_m)(\hat x^j_n -\hat  x^j_m)}
\end{align}
{is the generalized distance determined from the permittivity tensor.} These equalities are valid {in the presence of anisotropy} since the first P-K stress is always determinable for hyper-elastic systems. The Hamiltonian  no longer depends on the metric perturbations $\delta g_{ij} = \Lambda_i^k\eta'_{jk} + \Lambda_j^k \eta'_{ik}$, {therefore the second P-K and Cauchy stress must be defined by equation \eqref{eq:SecondPiola} and the first equality in equation \eqref{eq:Cauchy} respectively.}
Whereas the second P-K and Cauchy stress are definable in the absence of $\delta g$ as a perturbation, the second elasticity is not as it by definition is the variation of $\mathcal T$ by the metric. 

{The diamagnetic term for the first stress is calculated using equation \eqref{eq:deltaTresult} from the main text.}
\begin{align}
    (\delta _\Lambda \hat T)^{i~k}_{~j~l} &= - i[\hat{J}^k_l,(\hat T_0)^i_j] - \delta^i_l (\hat T_0)^k_j\\
    &= -\frac{1}{2}\sum_{n} \left( (\tilde m^{-1})^{ik}  \{\hat \pi_l^n,\hat \pi_j^n\}  - \delta ^k_j  (\widetilde m^{-1})^{ip}  \{\hat \pi_p^n,\hat \pi_l^n\}  - \delta^i_l (\widetilde m^{-1})^{kp}  \{\hat \pi_p^n,\hat \pi_j^n\}\right) \\ &  - \frac{1}{2}\sum_{n \neq m} \bigg( V''(|\hat {\textbf{x}}_n - \hat {\textbf{x}}_m|_\epsilon) - \frac{V'(|\hat {\textbf{x}}_n - \hat {\textbf{x}}_m|_\epsilon)}{|\hat {\textbf{x}}_n -\hat {\textbf{x}}_m|_\epsilon }\bigg) \frac{(\hat x^i_n - \hat x^i_m)\epsilon
    _{jq}(\hat x^q_n - \hat x^q_m)}{|\hat {\textbf{x}}_n - \hat {\textbf{x}}_m|_\epsilon} \frac{(\hat x^k_n - \hat x^k_m)\epsilon_{wl}(\hat x^w_n - \hat x^w_m)}{|\hat {\textbf{x}}_n - \hat {\textbf{x}}_m|_\epsilon} \\
   &-   \frac{1}{2}\sum_{n \neq m} V'(|\hat {\textbf{x}}_n - \hat {\textbf{x}}_m|_\epsilon)\frac{(\hat x^k_n - \hat x^k_m)\epsilon_{jl} \left(   \hat x^i_n - \hat x^i_m   \right)}{|\hat {\textbf{x}}_n - \hat {\textbf{x}}_m|_\epsilon}
\end{align}

\section{Calculation of $\delta_\Lambda \hat T$}
\label{sec:A0}

{In this appendix we derive Eq.~}\eqref{eq:deltaTresult}.
Starting with equations \eqref{eq:perturbation}, \eqref{eq:Hprime}, the identity $\lambda = \log \Lambda$, and the definition of $\delta _\Lambda \hat T$ in equation \eqref{eq:deltaT} we have
\begin{align}
   (\delta_\Lambda \hat T)_{~m~p}^{n~q} = \frac{\partial^2 }{\partial \Lambda_n^m\partial \Lambda_q^p} \bigg(  i (\log\Lambda)_i^j [\hat{J}^i_j, \hat H] + \frac{1}{2}(\log\Lambda)_{i}^j (\log\Lambda)_k^l[\hat{J}^i_j ,[J_{l}^k,\hat H]]+...\bigg)  .
\end{align}
{Expanding} to quadratic order in $\delta \Lambda = \Lambda - \mathbb 1$ yields
\begin{align}
    (\delta_\Lambda \hat T)_{~m~p}^{n~q}=   \frac{\partial^2 }{\partial \Lambda_n^m\partial \Lambda_q^p} \bigg( i \delta\Lambda_i^j [\hat{J}^i_j, \hat H] + \frac{1}{2} \delta \Lambda_{i}^j \delta \Lambda_k^l[\hat{J}^i_j ,[J_{l}^k,\hat H]] - \frac{i}{2}\delta \Lambda_i^k \delta \Lambda_{k}^j [\hat{J}^i_j,\hat H]\bigg) .
\end{align}
{To proceed further, take the derivative and simplify using the Ward identity [Eq. \eqref{eq:Ward}] to find}
\begin{align}
    (\delta_\Lambda \hat T)_{~m~p}^{n~q}&= \frac{1}{2}[\hat J_m^n,[\hat J_p^q,\hat H]] + \frac{1}{2}[\hat J_p^q,[\hat J_m^n,\hat H]] - \frac{i}{2}\delta_m^q [\hat J^n_p, \hat H] -\frac{i}{2}\delta_p^n [\hat J_m^q,\hat H],\\
    &= -\frac{i}{2}[\hat J_m^n,(\hat T_0)_p^q] - \frac{i}{2}[\hat J_p^q,(\hat T_0)_m^n] - \frac{1}{2}\delta_m^q (\hat T_0)^n_p -\frac{1}{2}\delta_p^n (\hat T_0)_m^q.\label{eq:deltaTresult2}
\end{align}
Observe that the proper second derivative in equation \eqref{eq:deltaTresult} is not simply the symmetrization of $\Delta _\lambda\hat T_0$, but also contains terms from $\Gamma$ in equation \eqref{eq:ChainRule}.
Note that equation \eqref{eq:deltaTresult2} is manifestly symmetric about the hyper-elastic exchange $(nm)\leftrightarrow(qp)$.
We may simplify Eq.~\eqref{eq:deltaTresult2} using a combination of the Ward identity [Eq. \eqref{eq:Ward}] and the Jacobi identity [Eq. \eqref{eq:Jacobi}] to arrive at
\begin{align}
    - \frac{i}{2}[\hat{J}^i_j,(\hat T_0)^k_l ] = \frac{1}{2}\delta^k_j (\hat T_0)^i_l - \frac{1}{2}\delta^i_l (\hat T_0)^k_j - \frac{i}{2}[\hat{J}^k_l,(\hat T_0)^i_j],
\end{align}
implying
\begin{align}
(\delta_\Lambda\hat T)_{~m~p}^{n~q} = i[(\hat T_0)^n_m,J^q_p] - \delta^n_p (\hat T_0)^q_m,
\end{align}
which is the equation \eqref{eq:deltaTresult} used in the main text.

\twocolumngrid

\bibliography{refs.bib}

@book{WeinbergVol2,
    author = "Weinberg, Steven",
    title = "{The quantum theory of fields. Vol. 2: Modern applications}",
    doi = "10.1017/CBO9781139644174",
    isbn = "978-1-139-63247-8, 978-0-521-67054-8, 978-0-521-55002-4",
    publisher = "Cambridge University Press",
    month = "8",
    year = "2013"
}

@article{dashen1969chiral,
  title = {Chiral $\mathrm{SU}(3)\ensuremath{\bigotimes}\mathrm{SU}(3)$ as a Symmetry of the Strong Interactions},
  author = {Dashen, Roger},
  journal = {Phys. Rev.},
  volume = {183},
  issue = {5},
  pages = {1245--1260},
  numpages = {0},
  year = {1969},
  month = {Jul},
  publisher = {American Physical Society},
  doi = {10.1103/PhysRev.183.1245},
  url = {https://link.aps.org/doi/10.1103/PhysRev.183.1245}
}

@article{abanov2019freesurface,
  title = {Free-{{Surface Variational Principle}} for an {{Incompressible Fluid}} with {{Odd Viscosity}}},
  author = {Abanov, Alexander G and Monteiro, Gustavo M},
  year = {2019},
  journal = {Physical review letters},
  volume = {122},
  number = {15},
  pages = {154501},
  publisher = {{APS}}
}

@article{abanov2020hydrodynamics,
  title = {Hydrodynamics of Two-Dimensional Compressible Fluid with Broken Parity: {{Variational}} Principle and Free Surface Dynamics in the Absence of Dissipation},
  author = {Abanov, Alexander G and Can, Tankut and Ganeshan, Sriram and Monteiro, Gustavo M},
  year = {2020},
  journal = {Physical Review Fluids},
  volume = {5},
  number = {10},
  pages = {104802},
  publisher = {{APS}}
}

@article{avron1987adiabatic,
  title = {Adiabatic Theorems and Applications to the Quantum {{Hall}} Effect},
  author = {Avron, {\relax JE} and Seiler, Rued and Yaffe, {\relax LG}},
  year = {1987},
  journal = {Communications in Mathematical Physics},
  volume = {110},
  number = {1},
  pages = {33--49},
  publisher = {{Springer}}
}

@article{avron1995viscosity,
  title = {Viscosity of {{Quantum Hall Fluids}}},
  author = {Avron, J. E. and Seiler, R. and Zograf, P. G.},
  year = {1995},
  month = jan,
  journal = {Phys Rev Lett},
  volume = {75},
  number = {4},
  pages = {697--700},
  url = {http://www.google.com/search?client=safari&rls=en-us&q=VISCOSITY+OF+QUANTUM+HALL+FLUIDS&ie=UTF-8&oe=UTF-8},
  langid = {english},
  pmid = {A1995RK69000026}
}

@article{berdyugin2019measuring,
  title = {Measuring {{Hall}} Viscosity of Graphene's Electron Fluid},
  author = {Berdyugin, Alexey I and Xu, {\relax SG} and Pellegrino, {\relax FMD} and Kumar, R Krishna and Principi, Alessandro and Torre, Iacopo and Shalom, M Ben and Taniguchi, Takashi and Watanabe, Kenji and Grigorieva, {\relax IV} and others},
  year = {2019},
  journal = {Science},
  volume = {364},
  number = {6436},
  pages = {162--165},
  publisher = {{American Association for the Advancement of Science}}
}

@article{bradlyn2012kubo,
  title = {Kubo Formulas for Viscosity: {{Hall}} Viscosity, {{Ward}} Identities, and the Relation with Conductivity},
  author = {Bradlyn, Barry and Goldstein, Moshe and Read, N},
  year = {2012},
  journal = {Physical Review B},
  volume = {86},
  number = {24},
  pages = {245309},
  publisher = {{APS}}
}

@article{cooper1997thermoelectric,
  title = {Thermoelectric Response of an Interacting Two-Dimensional Electron Gas in a Quantizing Magnetic Field},
  author = {Cooper, {\relax NR} and Halperin, {\relax BI} and Ruzin, {\relax IM}},
  year = {1997},
  journal = {Physical Review B},
  volume = {55},
  number = {4},
  pages = {2344},
  publisher = {{APS}}
}

@inproceedings{forster1975hydrodynamic,
  title = {Hydrodynamic Fluctuations, Broken Symmetry, and Correlation Functions},
  booktitle = {Reading, {{Mass}}., {{WA Benjamin}}, {{Inc}}.({{Frontiers}} in {{Physics}}. {{Volume}} 47), 1975. 343 p.},
  author = {Forster, Dieter},
  year = {1975},
  volume = {47}
}

@article{irving1950statistical,
  title = {The Statistical Mechanical Theory of Transport Processes. {{IV}}. {{The}} Equations of Hydrodynamics},
  author = {Irving, {\relax JH} and Kirkwood, John G},
  year = {1950},
  journal = {The Journal of chemical physics},
  volume = {18},
  number = {6},
  pages = {817--829},
  publisher = {{AIP}}
}

@article{levay1995berry,
  title = {Berry Phases for {{Landau Hamiltonians}} on Deformed Tori},
  author = {L{\'e}vay, P{\'e}ter},
  year = {1995},
  journal = {J. Math. Phys.},
  volume = {36},
  number = {6},
  pages = {2792},
  doi = {10.1063/1.531066}
}

@article{link2018elastic,
  title = {Elastic Response of the Electron Fluid in Intrinsic Graphene: {{The}} Collisionless Regime},
  author = {Link, Julia M and Sheehy, Daniel E and Narozhny, Boris N and Schmalian, J{\"o}rg},
  year = {2018},
  journal = {Physical Review B},
  volume = {98},
  number = {19},
  pages = {195103},
  publisher = {{APS}}
}

@article{lucas2018hydrodynamics,
  title = {Hydrodynamics of Electrons in Graphene},
  author = {Lucas, Andrew and Fong, Kin Chung},
  year = {2018},
  journal = {Journal of Physics: Condensed Matter},
  volume = {30},
  number = {5},
  pages = {053001},
  publisher = {{IOP Publishing}}
}

@article{luttinger1964theory,
  title = {Theory of {{Thermal Transport Coefficients}}},
  author = {Luttinger, J. M.},
  year = {1964},
  month = sep,
  journal = {Phys. Rev.},
  volume = {135},
  number = {6A},
  pages = {A1505--A1514},
  publisher = {{American Physical Society}},
  doi = {10.1103/PhysRev.135.A1505},
  url = {https://link.aps.org/doi/10.1103/PhysRev.135.A1505}
}

@article{moll2016evidence,
  title = {Evidence for Hydrodynamic Electron Flow in {{PdCoO2}}},
  author = {Moll, Philip JW and Kushwaha, Pallavi and Nandi, Nabhanila and Schmidt, Burkhard and Mackenzie, Andrew P},
  year = {2016},
  journal = {Science},
  volume = {351},
  number = {6277},
  pages = {1061--1064},
  publisher = {{American Association for the Advancement of Science}}
}

@article{offertaler2019viscoelastic,
  title = {Viscoelastic Response of Quantum {{Hall}} Fluids in a Tilted Field},
  author = {Offertaler, Bendeguz and Bradlyn, Barry},
  year = {2019},
  journal = {Physical Review B},
  volume = {99},
  number = {3},
  pages = {035427},
  publisher = {{APS}}
}

@article{pellegrino2017nonlocal,
  title = {Nonlocal Transport and the {{Hall}} Viscosity of Two-Dimensional Hydrodynamic Electron Liquids},
  author = {Pellegrino, Francesco MD and Torre, Iacopo and Polini, Marco},
  year = {2017},
  journal = {Physical Review B},
  volume = {96},
  number = {19},
  pages = {195401},
  publisher = {{APS}}
}

@article{krishna2017superballistic,
  title={Superballistic flow of viscous electron fluid through graphene constrictions},
  author={Krishna Kumar, R and Bandurin, DA and Pellegrino, FMD and Cao, Y and Principi, A and Auton, GH and Ben Shalom, M and Falkovich, G and others},
  journal={Nature Physics},
  volume={13},
  number={12},
  pages={1182--1185},
  year={2017},
  publisher={Nature Publishing Group UK London}
}

@article{levitov2016electron,
  title={Electron viscosity, current vortices and negative nonlocal resistance in graphene},
  author={Levitov, Leonid and Falkovich, Gregory},
  journal={Nature Physics},
  volume={12},
  number={7},
  pages={672--676},
  year={2016},
  publisher={Nature Publishing Group UK London}
}

@article{bandurin2016negative,
  title={Negative local resistance caused by viscous electron backflow in graphene},
  author={Bandurin, DA and Kumar, R Krishna and Ben Shalom, M and Principi, A and Auton, GH and Khestanova, E and Novoselov, KS and Grigorieva, IV and others},
  journal={Science},
  volume={351},
  number={6277},
  pages={1055--1058},
  year={2016},
  publisher={American Association for the Advancement of Science}
}

@article{rao2020hall,
  title = {Hall {{Viscosity}} in {{Quantum Systems}} with {{Discrete Symmetry}}: {{Point Group}} and {{Lattice Anisotropy}}},
  author = {Rao, Pranav and Bradlyn, Barry},
  year = {2020},
  month = apr,
  journal = {Phys. Rev. X},
  volume = {10},
  number = {2},
  pages = {021005},
  publisher = {{American Physical Society}},
  doi = {10.1103/PhysRevX.10.021005},
  url = {https://link.aps.org/doi/10.1103/PhysRevX.10.021005}
}

@article{read2009nonabelian,
  title = {Non-{{Abelian}} Adiabatic Statistics and {{Hall}} Viscosity in Quantum {{Hall}} States and p {{X}}+ i p y Paired Superfluids},
  author = {Read, N},
  year = {2009},
  journal = {Physical Review B},
  volume = {79},
  number = {4},
  pages = {045308},
  publisher = {{APS}}
}

@article{read2011hall,
  title = {Hall Viscosity, Orbital Spin, and Geometry: {{Paired}} Superfluids and Quantum {{Hall}} Systems},
  author = {Read, N and Rezayi, {\relax EH}},
  year = {2011},
  journal = {Physical Review B},
  volume = {84},
  number = {8},
  pages = {085316},
  publisher = {{APS}}
}

@article{scheibner2019odd,
   title={Odd elasticity},
   volume={16},
   ISSN={1745-2481},
   url={http://dx.doi.org/10.1038/s41567-020-0795-y},
   DOI={10.1038/s41567-020-0795-y},
   number={4},
   journal={Nature Physics},
   publisher={Springer Science and Business Media LLC},
   author={Scheibner, Colin and Souslov, Anton and Banerjee, Debarghya and Surówka, Piotr and Irvine, William T. M. and Vitelli, Vincenzo},
   year={2020},
   month=mar, pages={475–480} }

@article{soni2019odd,
  title = {The Odd Free Surface Flows of a Colloidal Chiral Fluid},
  author = {Soni, Vishal and Bililign, Ephraim S and Magkiriadou, Sofia and Sacanna, Stefano and Bartolo, Denis and Shelley, Michael J and Irvine, William TM},
  year = {2019},
  journal = {Nature Physics},
  volume = {15},
  number = {11},
  pages = {1188--1194},
  publisher = {{Nature Publishing Group}}
}

@article{tokatly2006magnetoelasticity,
  title = {Magnetoelasticity Theory of Incompressible Quantum {{Hall}} Liquids},
  author = {Tokatly, I. V.},
  year = {2006},
  month = jan,
  journal = {Phys Rev B},
  volume = {73},
  number = {20},
  pages = {205340},
  doi = {10.1103/PhysRevB.73.205340},
  url = {http://prb.aps.org/abstract/PRB/v73/i20/e205340},
  langid = {english},
  pmid = {000237950500076}
}

@article{tokatly2007lorentz,
  title = {Lorentz Shear Modulus of a Two-Dimensional Electron Gas at High Magnetic Field},
  author = {Tokatly, I. V. and Vignale, G.},
  year = {2007},
  month = jan,
  journal = {Phys Rev B},
  volume = {76},
  number = {16},
  pages = {161305},
  doi = {10.1103/PhysRevB.76.161305},
  url = {http://scitation.aip.org/getabs/servlet/GetabsServlet?prog=normal&id=PRBMDO000076000016161305000001&idtype=cvips&gifs=yes},
  langid = {english},
  pmid = {000250620600011}
}

@article{critelli2014anisotropic,
  title={Anisotropic shear viscosity of a strongly coupled non-Abelian plasma from magnetic branes},
  author={Critelli, R and Finazzo, Stefano Ivo and Zaniboni, M and Noronha, J},
  journal={Physical Review D},
  volume={90},
  number={6},
  pages={066006},
  year={2014},
  publisher={APS}
}

@article{nair2021topological,
  title = {Topological terms and diffeomorphism anomalies in fluid dynamics and sigma models},
  author = {Nair, V. P.},
  journal = {Phys. Rev. D},
  volume = {103},
  issue = {8},
  pages = {085017},
  numpages = {11},
  year = {2021},
  month = {Apr},
  publisher = {American Physical Society},
  doi = {10.1103/PhysRevD.103.085017},
  url = {https://link.aps.org/doi/10.1103/PhysRevD.103.085017}
}

@article{tong2023gauge,
  title={A gauge theory for shallow water},
  author={Tong, David},
  journal={SciPost Physics},
  volume={14},
  number={5},
  pages={102},
  year={2023}
}

@article{monteiro2024korteweg,
  title={Korteweg de-Vries Dynamics at the Edge of Laughlin State},
  author={Monteiro, Gustavo M and Ganeshan, Sriram},
  journal={arXiv preprint arXiv:2410.01730},
  year={2024}
}

@article{monteiro2024topological,
  title={Topological fluids with boundaries and fractional quantum Hall edge dynamics: A fluid dynamics derivation of the chiral boson action},
  author={Monteiro, Gustavo M and Nair, VP and Ganeshan, Sriram},
  journal={Physical Review B},
  volume={109},
  number={17},
  pages={174525},
  year={2024},
  publisher={APS}
}

@article{machado2023hamiltonian,
  title={Hamiltonian structure of 2D fluid dynamics with broken parity},
  author={Machado Monteiro, Gustavo and Abanov, Alexander G and Ganeshan, Sriram},
  journal={SciPost Physics},
  volume={14},
  number={5},
  pages={103},
  year={2023}
}

@article{Ian2024,
  title = {Anisotropy of hydrostatic stress for Hall droplets with in-plane magnetic field},
  author = {Osborne, Ian and Monteiro, Gustavo M. and Bradlyn, Barry},
  journal = {Phys. Rev. B},
  volume = {110},
  issue = {11},
  pages = {115141},
  numpages = {14},
  year = {2024},
  month = {Sep},
  publisher = {American Physical Society},
  doi = {10.1103/PhysRevB.110.115141},
  url = {https://link.aps.org/doi/10.1103/PhysRevB.110.115141}
}

@book{landau_elasticity,
  author = {Landau, L. D. and Lifshitz, E. M.},
  title = {Theory of Elasticity},
  series = {Course of Theoretical Physics, Vol. 7},
  publisher = {Pergamon Press},
  address = {Oxford},
  year = {1970},
  edition = {3rd},
  note = {Translated from Russian by J. B. Sykes and W. H. Reid}
}

@book{Marsden_Elasticity,
  title={Mathematical Foundations of Elasticity},
  author={Marsden, J.E. and Hughes, T.J.R.},
  isbn={9780486142272},
  series={Dover Civil and Mechanical Engineering},
  url={https://books.google.com/books?id=-STEAgAAQBAJ},
  year={2012},
  publisher={Dover Publications}
}

@misc{folkner2024elasticmodulithermalconductivity,
      title={Elastic moduli and thermal conductivity of quantum materials at finite temperature}, 
      author={Dylan A. Folkner and Zekun Chen and Giuseppe Barbalinardo and Florian Knoop and Davide Donadio},
      year={2024},
      eprint={2409.09551},
      archivePrefix={arXiv},
      primaryClass={cond-mat.mtrl-sci},
      url={https://arxiv.org/abs/2409.09551}, 
}

@article{deJong2015,
  author       = {de Jong, Maarten and Chen, Wei and Angsten, Thomas and Jain, Anubhav and Notestine, Randy and Gamst, Anthony and Sluiter, Marcel and Ande, Chaitanya K. and van der Zwaag, Sybrand and Plata, José J. and Toher, Cormac and Curtarolo, Stefano and Ceder, Gerbrand and Persson, Kristin A. and Asta, Mark},
  title        = {Charting the complete elastic properties of inorganic crystalline compounds},
  journal      = {Scientific Data},
  volume       = {2},
  number       = {150009},
  year         = {2015},
  doi          = {10.1038/sdata.2015.9},
  url          = {https://doi.org/10.1038/sdata.2015.9}
}

@article{ElasticityPathIntegral2001,
  title = {Elastic constants of quantum solids by path integral simulations},
  author = {Sch\"offel, Philipp and M\"user, Martin H.},
  journal = {Phys. Rev. B},
  volume = {63},
  issue = {22},
  pages = {224108},
  numpages = {9},
  year = {2001},
  month = {May},
  publisher = {American Physical Society},
  doi = {10.1103/PhysRevB.63.224108},
  url = {https://link.aps.org/doi/10.1103/PhysRevB.63.224108}
}

@article{OddElasticity2024,
  title = {Odd elasticity and topological waves in active surfaces},
  author = {Fossati, Michele and Scheibner, Colin and Fruchart, Michel and Vitelli, Vincenzo},
  journal = {Phys. Rev. E},
  volume = {109},
  issue = {2},
  pages = {024608},
  numpages = {16},
  year = {2024},
  month = {Feb},
  publisher = {American Physical Society},
  doi = {10.1103/PhysRevE.109.024608},
  url = {https://link.aps.org/doi/10.1103/PhysRevE.109.024608}
}

@article{Randeria_Viscosity,
  title = {Viscosity of strongly interacting quantum fluids: Spectral functions and sum rules},
  author = {Taylor, Edward and Randeria, Mohit},
  journal = {Phys. Rev. A},
  volume = {81},
  issue = {5},
  pages = {053610},
  numpages = {16},
  year = {2010},
  month = {May},
  publisher = {American Physical Society},
  doi = {10.1103/PhysRevA.81.053610},
  url = {https://link.aps.org/doi/10.1103/PhysRevA.81.053610}
}

@article{Mendez_Valderrama_2024,
   title={Low-Energy Optical Sum Rule in Moiré Graphene},
   volume={133},
   ISSN={1079-7114},
   url={http://dx.doi.org/10.1103/PhysRevLett.133.196501},
   DOI={10.1103/physrevlett.133.196501},
   number={19},
   journal={Physical Review Letters},
   publisher={American Physical Society (APS)},
   author={Mendez-Valderrama, J. F. and Mao, Dan and Chowdhury, Debanjan},
   year={2024},
   month=nov }

@article{mao2025lowenergy,
  title = {Low-Energy Optical Absorption in Correlated Insulators: {{Projected}} Sum Rules and the Role of Quantum Geometry},
  shorttitle = {Low-Energy Optical Absorption in Correlated Insulators},
  author = {Mao, Dan and {Mendez-Valderrama}, Juan Felipe and Chowdhury, Debanjan},
  year = 2025,
  month = aug,
  journal = {Physical Review B},
  volume = {112},
  number = {7},
  pages = {075116},
  publisher = {American Physical Society},
  doi = {10.1103/xmz7-jgl6},
  url = {https://link.aps.org/doi/10.1103/xmz7-jgl6},
  urldate = {2026-02-12}
}

@article{ROSTAMI2021168523,
title = {Gauge invariance and Ward identities in nonlinear response theory},
journal = {Annals of Physics},
volume = {431},
pages = {168523},
year = {2021},
issn = {0003-4916},
doi = {https://doi.org/10.1016/j.aop.2021.168523},
url = {https://www.sciencedirect.com/science/article/pii/S0003491621001299},
author = {Habib Rostami and Mikhail I. Katsnelson and Giovanni Vignale and Marco Polini}
}

@article{Brown_2021,
   title={Elasticity theory in general relativity},
   volume={38},
   ISSN={1361-6382},
   url={http://dx.doi.org/10.1088/1361-6382/abe1ff},
   DOI={10.1088/1361-6382/abe1ff},
   number={8},
   journal={Classical and Quantum Gravity},
   publisher={IOP Publishing},
   author={Brown, J David},
   year={2021},
   month=mar, pages={085017} }

@article{SecondOrderSumRules,
  title = {Spectral density and sum rules for second-order response functions},
  author = {Bradlyn, Barry and Abbamonte, Peter},
  journal = {Phys. Rev. B},
  volume = {110},
  issue = {24},
  pages = {245132},
  numpages = {15},
  year = {2024},
  month = {Dec},
  publisher = {American Physical Society},
  doi = {10.1103/PhysRevB.110.245132},
  url = {https://link.aps.org/doi/10.1103/PhysRevB.110.245132}
}

@article{mckay2024charge,
  title = {Charge Conservation beyond Uniformity: {{Spatially}} Inhomogeneous Electromagnetic Response in Periodic Solids},
  author = {McKay, Robert C and Mahmood, Fahad and Bradlyn, Barry},
  year = 2024,
  journal = {Physical Review X},
  volume = {14},
  number = {1},
  pages = {011058},
  publisher = {APS},
  doi = {10.1103/PhysRevX.14.011058},
  url = {https://link.aps.org/doi/10.1103/PhysRevX.14.011058}
}

@article{bradlyn2024spectral,
  title = {Spectral Density and Sum Rules for Second-Order Response Functions},
  author = {Bradlyn, Barry and Abbamonte, Peter},
  year = 2024,
  journal = {arXiv preprint arXiv:2404.16144},
  eprint = {2404.16144},
  doi = {10.48550/arXiv.2404.16144},
  url = {https://doi.org/10.48550/arXiv.2404.16144},
  archiveprefix = {arXiv}
}

@article{watanabe2020generalized,
  title = {Generalized \$f\$-Sum Rules and {{Kohn}} Formulas on Nonlinear Conductivities},
  author = {Watanabe, Haruki and Oshikawa, Masaki},
  year = 2020,
  month = oct,
  journal = {Physical Review B},
  volume = {102},
  number = {16},
  pages = {165137},
  publisher = {American Physical Society},
  doi = {10.1103/PhysRevB.102.165137},
  url = {https://link.aps.org/doi/10.1103/PhysRevB.102.165137},
  urldate = {2023-10-26}
}

@article{Hughes_CSMM,
  title = {Hall viscosity and geometric response in the Chern-Simons matrix model of the Laughlin states},
  author = {Lapa, Matthew F. and Hughes, Taylor L.},
  journal = {Phys. Rev. B},
  volume = {97},
  issue = {20},
  pages = {205122},
  numpages = {25},
  year = {2018},
  month = {May},
  publisher = {American Physical Society},
  doi = {10.1103/PhysRevB.97.205122},
  url = {https://link.aps.org/doi/10.1103/PhysRevB.97.205122}
}

@article{Kazemirad2013,
  author    = {Kazemirad, Siavash and Heris, Hossein K. and Mongeau, Luc},
  title     = {Experimental methods for the characterization of the frequency-dependent viscoelastic properties of soft materials},
  journal   = {The Journal of the Acoustical Society of America},
  shortjournal = {J. Acoust. Soc. Am.},
  year      = {2013},
  volume    = {133},
  number    = {5},
  pages     = {3186--3197},
  month     = may,
  doi       = {10.1121/1.4798668},
  pmid      = {23654420},
  pmcid     = {PMC3663851},
  publisher = {Acoustical Society of America},
  address   = {United States},
}

@article{LORENZ2014565,
title = {Master curve of viscoelastic solid: Using causality to determine the optimal shifting procedure, and to test the accuracy of measured data},
journal = {Polymer},
volume = {55},
number = {2},
pages = {565-571},
year = {2014},
issn = {0032-3861},
doi = {https://doi.org/10.1016/j.polymer.2013.12.033},
url = {https://www.sciencedirect.com/science/article/pii/S0032386113011336},
author = {B. Lorenz and W. Pyckhout-Hintzen and B.N.J. Persson}
}

@misc{jain2025accoustic,
      title={Topological Acoustic Diode}, 
      author={Ashwat Jain and Wojciech J. Jankowski and M. Mehraeen and Robert-Jan Slager},
      year={2026},
      eprint={2601.20951},
      archivePrefix={arXiv},
      primaryClass={cond-mat.mes-hall},
      url={https://arxiv.org/abs/2601.20951}, 
}

@misc{jain2025nonlinearoddviscoelasticeffect,
      title={Nonlinear Odd Viscoelastic Effect}, 
      author={Ashwat Jain and Wojciech J. Jankowski and M. Mehraeen and Robert-Jan Slager},
      year={2025},
      eprint={2511.22706},
      archivePrefix={arXiv},
      primaryClass={cond-mat.mes-hall},
      url={https://arxiv.org/abs/2511.22706}, 
}

\end{document}